\documentstyle[11pt,amssymb,amsfonts,amsthm,amssymb,amscd,amstext,epsfig]{article}

\input amssym.def
\input amssym.tex
\def\ben{\begin{equation}}
\def\beq{\begin{equation}}
\def\een{\end{equation}}
\def\eeq{\end{equation}}

\let\a=\alpha \let\b=\beta \let\g=\gamma  
\let\z=\zeta  \let\q=\theta 
    \let\p=\phi 
\let\s=\sigma \let\t=\tau  

\let\w=\omega \let\G=\Gamma

\let\pa=\partial
\def\be{\begin{equation}}
\def\ee{\end{equation}}
\def\ba{\begin{array}}
\def\ea{\end{array}}

\def\dalemb#1#2{{\vbox{\hrule height .#2pt
        \hbox{\vrule width.#2pt height#1pt \kern#1pt
                \vrule width.#2pt}
        \hrule height.#2pt}}}
\def\square{\mathord{\dalemb{6.8}{7}\hbox{\hskip1pt}}}
\newcommand{\bea}{\begin{eqnarray}}
\newcommand{\eea}{\end{eqnarray}}

\newcommand{\Tr}{{\rm Tr} }
\newcommand{\beqa}{\begin{eqnarray}}
\newcommand{\eeqa}{\end{eqnarray}}

\def\R{{{\Bbb R}}}
\def\Z{{{\Bbb Z}}}

\def\Z{{{\Bbb Z}}}

\def\Re{{{\frak{Re}}}}

\begin{document}

\begin{flushright}
DAMTP-2004-73\\
hep-th/0407030
\end{flushright}

\begin{center}

\vspace{2cm}

{\LARGE {\bf Properties of some five dimensional Einstein metrics} }

\vspace{1cm}

Gary W. Gibbons$^{\sharp}$ , Sean A. Hartnoll$^{\sharp}$ and Yukinori Yasui$^{\flat}$

\vspace{1cm}

$^\sharp$ {\it DAMTP, Centre for Mathematical Sciences,
 Cambridge University\\ Wilberforce Road, Cambridge CB3 OWA, UK}

\vspace{0.3cm}

$^\flat$ {\it Department of Physics, Osaka City University \\
  Sumiyoshi-ku, Osaka, 558-8585, Japan }

\end{center}

\vspace{1cm}

\begin{abstract}

The volumes, spectra and geodesics of a recently
constructed infinite family of five-dimensional inhomogeneous
Einstein metrics on the two $S^3$ bundles over $S^2$ are
examined. The metrics are in general of cohomogeneity one
but they contain the infinite family of homogeneous metrics $T^{p,1}$.
The geodesic flow is shown to be completely integrable, in fact
both the Hamilton-Jacobi and the Laplace equation separate. As an
application of these results, we compute the zeta function of the
Laplace operator on $T^{p,1}$ for large $p$.
We discuss the spectrum of the Lichnerowicz operator on symmetric
transverse tracefree second rank tensor fields, with application to
the stability of Freund-Rubin compactifications and generalised black
holes.

\end{abstract}

\pagebreak
\setcounter{page}{1}

\tableofcontents

\setcounter{equation}{0}

\vspace{.5truecm}

\section{Introduction}

Compact Riemannian Einstein manifolds may be used as basic building blocks for
solutions to higher dimensional gravity and supergravity theories. An
important example in recent years has been Freund-Rubin compactification
in the context of the AdS/CFT correspondence. Here one has backgrounds
such as $AdS_5\times M^5$ supported by 5-form flux. The Einstein
manifold $M$ encodes geometrically the properties of the dual conformal
field theory such as the R-symmetry and central charge
\cite{Gubser:1998vd}. Another set of examples are
generalised $D$-dimensional black holes, where the horizon is given by
an arbitrary Einstein manifold $M$ rather than the usual sphere
$S^{D-2}$\cite{Gibbons:2002pq} . In fact, the two examples we have just given are
related. A generalised cone spacetime over the Einstein manifold $M$ is a Ricci
flat Lorentzian metric
\be
ds^2 = -dt^2 + dr^2 + r^2 ds^2_M \,.
\ee
Generalised black holes may be thought of as black holes in generalised
cone spacetimes, see (\ref{eq:bh}) below. Instead of black holes, we could have
appended three more flat
directions and considered an extremal D3-brane sitting at the tip of
the cone. The near horizon geometry of this D3 brane is then $AdS_5\times
M^5$ \cite{Klebanov:1998hh}.

Homogeneous Einstein manifolds have been known for some time,
well-known examples in the physics literature include the round spheres $S^d$
and the five dimensional $T^{p,q}$ spaces.
It is harder to find explicit inhomogeneous Einstein metrics on compact
manifolds. When a metric is found, it is then useful to study its
properties, both to achieve a better geometric understanding of the
manifold and with a view to physical applications. One key question
for inhomogeneous manifolds is the separability of partial differential
equations such as the Laplace equation and the Hamilton-Jacobi
equation for geodesics. Separability is an important first step
for being able to perform calculations involving the metrics.

In four dimensions, the first explicit inhomogeneous compact Einstein metric
was constructed by Page \cite{Page:zv}. Topologically the manifold is the
nontrivial $S^2$ bundle over $S^2$, which is isomorphic to
${\mathbb{CP}}^2\#\overline{{\mathbb{CP}}}^2$. The method of \cite{Page:zv} was
generalised recently to obtain, amongst other results, an infinite
series of inhomogeneous Einstein metrics in five dimensions on
$S^2\times S^3$ and on the nontrivial $S^3$ bundle over $S^2$
\cite{Hashimoto:2004kc}. The infinite series is parameterised by two integers
$(k_1,k_2)$. When $k_1=k_2\equiv k$, the metrics become the series of
homogeneous metrics $T^{k,1}$. This construction was then further
generalised to higher dimensions in \cite{Lu:2004ya}. In this paper we shall be
concerned with the properties of the five dimensional metrics labelled
by $(k_1,k_2)$ \cite{Hashimoto:2004kc}.

There have been two other recent explicit contructions of
inhomogeneous Einstein manifolds. Firstly, B\"ohm has constructed an infinite
family of inhomogeneous metrics on $S^5 \cdots S^9$ and products
of spheres \cite{bohm}. These metrics can be unwieldy
because the metric functions are not given explicitly, but as
solutions to nonlinear Ordinary Differential Equations. However, some
properties are
known\cite{Gibbons:2002th}. Secondly, an explicit infinite family of
inhomogeneous Einstein-Sasaki
manifolds were recently constructed on $S^2\times S^3$ in five
dimensions \cite{Gauntlett:2004yd}
and also generalised to higher dimensions \cite{Gauntlett:2004hh}.

An important question concerning Freund-Rubin compactifications and
generalised black hole spacetimes is whether they are stable. In both
cases, the question of stability reduces to the question of whether
the spectrum of the Lichnerowicz operator acting on rank two symmetric
tensor fields satisfies a certain lower bound
\cite{DeWolfe:2001nz,Gibbons:2002pq,Gibbons:2002th,Hartnoll:2003as}.
For the AdS/CFT correspondence one is more interested in stable
spacetimes, as this is when one expects a valid duality. For
generalised black holes spacetimes, one may also be interested in
unstable spacetimes because the endpoint of the instability is a
nontrivial and interesting dynamical question in higher dimensional
relativity. Finally, one may also consider the stability of
generalised black holes with a negative cosmological constant
\cite{Hartnoll:2003as}. In this case, the stability or instability of
the spacetime is related to a poorly understood phase transition in a
dual thermal field theory induced by the inhomogeneity of the background
\cite{Hartnoll:2003as,Hartnoll:2004kz}.

\subsection{Outline of the paper}

The organisation of this paper is as follows.

In section 2 below, we review the Einstein metrics constructed in
\cite{Hashimoto:2004kc}. We give a broad picture of the discrete moduli space
of metrics and we write the metrics in a form where the $SU(2)\times
U(1)$ isometry and the cohomogenity one property is manifest. We then go on
to characterise the moduli space in terms of volumes and Weyl
curvature eigenvalues. The latter allows us to comment on the
stability and instability of the resulting spacetimes.

In section 3 we study the Laplacian spectrum on the manifolds. In the
homogeneous cases, $T^{p,1}$, we give the spectrum explicitly. The spectrum on
$T^{p,q}$ has previously been studied in \cite{Gubser:1998vd}. In the
inhomogeneous case we show that the Laplace equation separates and for
slightly inhomogeneous metrics we give the spectrum as a perturbation
about the homogeneous cases.

In section 4 we use the results on the Laplacian spectrum to calculate
the zeta function on $T^{p,1}$ for large $p$. The zeta function
contains information such as the thermodynamics of a scalar gas on the
Einstein background. Ultimately, one would like to calculate the zeta
function for an inhomogeneous background, as the corresponding
thermodynamics may provide some insight into the phase transition
expected on backgrounds with a region of large curvature
\cite{Hartnoll:2004kz}.

In section 5 we study the geodesics of the Einstein metrics. We show
that the Hamilton-Jacobi equation separates and describe some
qualitative features of particle motion. For the homogeneous metrics we
give a full action-angle analysis of the geodesics. This allows us to
calculate the periods of geodesics and the semiclassical energy spectrum. The
semicalssical spectrum agrees with the full spectrum of the
Laplacian up to ordering ambiguities in quantisation.
In the inhomogeneous case the action-angle problem reduces
to an evaluation of elliptic integrals. These may be calculated in a
perturbation about the homogeneous metrics, and again the
semiclassical energy spectrum has a good agreement with the full Laplacian
spectrum.

Section 6 is the conclusion and contains suggestions for future research.

\section{The metrics}

The complete Einstein metrics presented in \cite{Hashimoto:2004kc}
depend upon two integers $k_1$ and $k_2$ which are the Chern numbers of
a principle $T^2$ bundle over $S^2$. The 5-manifold is then an
associated $S^3$ bundle over $S^2$. Topologically there are
two such bundles because $\pi_1(SO(4)) = \Z_2$. If $k_1+k_2$ is even
the bundle is trivial and if $k_1+k_2$ is odd the bundle is non-trivial.

The metrics may be written as
\be
ds^2_5 = h(\q)^2 d\q^2 + b(\q)^2 \left(d\chi^2+\sin^2\chi d\eta^2
\right) + a_{ij}(\q) (d\psi^i + \cos\chi d\eta) (d\psi^j + \cos\chi
d\eta)\,,
\ee
where the ranges of the angles are $0 \le \q \le \pi/2$, $0 \le
\chi \le \pi$, $0 \le \eta < 2\pi$ and $0 \le \psi^i <
4\pi/|k_i|$.

The general expressions for the metric quantities $h,b,a_{ij}$ and
the integers $k_i$ are given in appendix A. Throughout we will
use a normalisation such that the Ricci scalar of the metric is
$R= 20$. The explicit formulae for the metric depend upon two
constants $\nu_1$ and $\nu_2$ which depend in a complicated implicit
fashion (given in appendix A)
on the integers $k_1$ and $k_2$. We shall be concerned with the case
when $\nu_1,\nu_2 > 1$. In this case it appears that for
each pair of positive integers $k_1, k_2$ there is a unique pair
$\nu_1,\nu_2 > 1$. Note that there is a symmetry interchanging $k_1$
and $k_2$ and with it $\nu_1$ and $\nu_2$.

We will shortly consider the moduli space of Einstien metrics in some
detail. Let us first exhibit some special cases. A schematic summary
of the following statements is contained in Figure 1 below.

\medskip \noindent$\bullet$ If $k_1=k_2=k$, hence $\nu_1 = \nu_2 =
\nu$, we obtain the homogeneous
metrics known in the physics literature as $T^{k,1}$. After the change
of variables \cite{Hashimoto:2004kc}
\be
\b = 2\q \,, \qquad \g = \frac{1}{2} (\psi^2 - \psi^1) \,, \qquad t =
\frac{1}{2} (\psi^1 + \psi^2) \,,
\ee
the metrics takes the standard form for $T^{k,1}$
\bea
ds^2_{T^{k,1}} & = & \frac{1+\nu^2}{4(2+\nu^2)} \left(d\b^2 + \sin^2\b
d\g^2 \right) + \frac{1+\nu^2}{4(2\nu^2+1)} \left(d\chi^2 + \sin^2\chi
d\eta^2 \right) \nonumber \\
 & + & \frac{1+\nu^2}{2 (2+\nu^2)^2} \left( dt + \cos\b
d\g + k \cos\chi d\eta \right)^2\,,
\eea
where $k = \nu (\nu^2+2)/(2\nu^2+1)$.
The case $(k_1,k_2) = (1,1)$ is the Einstein-Sasaki metric known as $T^{1,1}$.

\medskip \noindent$\bullet$ The cases $(k_1,k_2) = (0,2)$ or $(2,0)$
have $\nu_1=1$ or $\nu_2=1$ and $\nu_2$ or $\nu_1$ arbitrary,
respectively. These are on the boundary of the cases we consider.
The metric is independent of $\nu_2$ or $\nu_1$ and coincides with the
round metric on  $S^5/\Z_2$.

\medskip \noindent $\bullet$. We can let $\nu_2 \rightarrow \infty$
with $\nu_1$ finite. One finds that
\be
\nu_1 \rightarrow { k_1 \over 2} + \sqrt { {k_1^2 \over 4} -1}\,, \qquad
  k_2 \rightarrow 0.
\ee
We call this the vertical limit.  The metric approaches the round
metric on $S^5/{\Bbb Z}_{k_1}$. It has an orbifold singularity along a
circle. This may be described as follows. The metric takes the form
\be
ds^2_{\infty} = d\theta ^2 + \sin ^2 \theta \,ds _3 ^2  + \cos^2  \theta d\psi^2  \,,
\ee
where $\psi$ has range $2\pi$ and
\be
ds_3^2 = \frac{1}{4} \left[ ( d \psi_1+ \cos \chi d \eta )^2 + d \chi ^2 +
\sin ^2 \chi d \eta ^2  \right] \,.
\ee
The angle $\psi_1$ is identified modulo $\frac{4
  \pi}{k_1}$. Therefore, there is an orbifold singularity along the
circle at $\theta =0$. Locally the singularity is $\R^4 /
\Z_{k_1}\times S^1$.

A word of caution about the vertical limit is necessary. Given that $k_2 \to 0$ and
$k_2$ is an integer, in fact the only solution is $k_2=0$. Thus,
although the limiting orbifold metrics we have just described exist,
there are no Einstein metrics `near' the limiting metric. A more
interesting limit is the limit in which we allow both $\nu_1$ and
$\nu_2$ to become large.

\medskip \noindent $\bullet$  For large values of $\nu_1$ and $\nu_2$ we find
that
\be
\nu_1  \approx {1 \over k_1} (k_1^2 + k_2^2 )\,,\qquad \nu_2  \approx
{1 \over k_ 2 } (k_1^2 + k_2^2)\,.
\ee
We call the limit in which both $\nu_1$ and $\nu_2$
become large, with the ratio $\nu_2/\nu_1 \sim k_1/k_2\equiv q$ fixed, the rational limit.
Near the rational limit, the metric becomes
\beq
ds^2\approx d\theta^2+\frac{\cos^2 \theta+q^2 \sin^2
\theta}{4(1+q^2)} ds^2_{S^2}+
\frac{\nu_2^2 q^2 \sin^2\q \cos^2\q (\w_1 - \w_2)^2 + (1+q^2) (q^4
  \sin^4\q\w_1^2 + \cos^4\q \w_2^2)}{4(1+q^2)^2 (\cos^2 \theta
  +q^2 \sin^2 \theta)}
 \,,
\eeq
where $\w_i = d\psi^i + \cos\chi d\eta$.
The actual limiting metric itself is not necessarily Einstein in this
limit. However, we do have an infinite sequence of Einstein metrics as
we approach the limit. Therefore, this limit is a richer source of
Einstein metrics than the vertical limit.
\begin{figure}[h]
\begin{center}
\epsfig{file=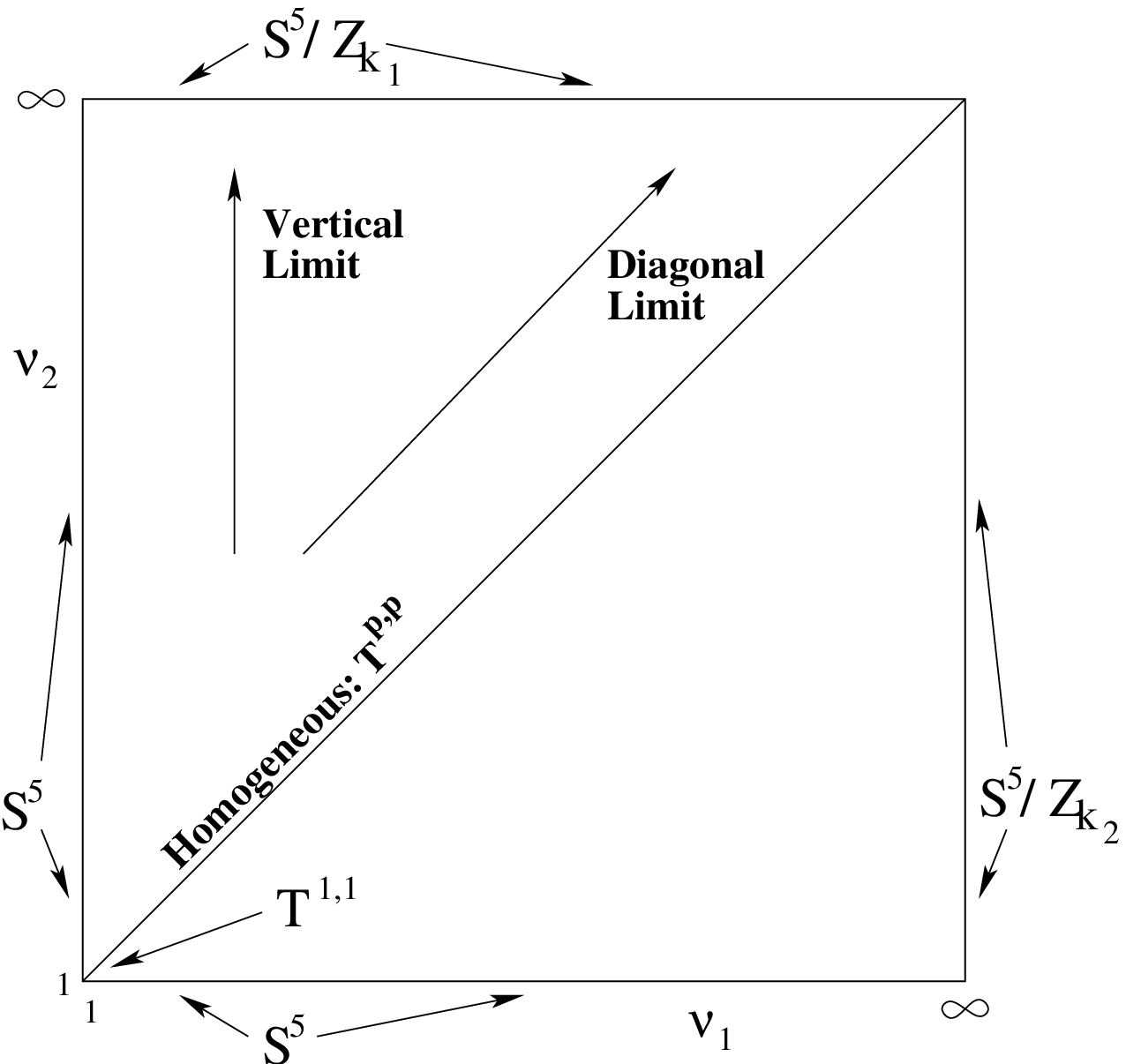,width=7cm}
\end{center}
\noindent {\bf Figure 1:} Schematic representation of the limits in
the moduli space of Einstein metrics. The figure is a little
misleading in that the rational limit should tend to the upper right
hand corner, and there are no metrics near the vertical limit.
\end{figure}

\subsection{The isometry group}

An $SU(2)_L\times U(1)_L \times U(1)_R$ isometry of the metric may
be made manifest by rewriting the metric in terms of
left-invariant $SU(2)\times U(1)$ forms
\bea
\s_1 & = & \cos\left(\frac{\psi^1+\psi^2}{2}\right) d\chi + \sin\left(\frac{\psi^1+\psi^2}{2}\right)
\sin\chi d\eta \,, \nonumber \\
\s_2 & = & -\sin\left(\frac{\psi^1+\psi^2}{2}\right) d\chi + \cos\left(\frac{\psi^1+\psi^2}{2}\right)
\sin\chi d\eta \,, \nonumber \\
\s_3 & = & \frac{1}{2} \left( d\psi_1 + d\psi_2 \right) + \cos\chi d\eta \,,
\nonumber \\
\s_4 & = & \frac{1}{2} \left( d\psi_1 - d\psi_2 \right) \,.
\eea
The metric becomes
\be
ds^2_5 = h^2 d\q^2 + b^2 \left(\s_1^2 + \s_2^2 \right) +
a_{11} (\s_3 + \s_4)^2 + a_{22} (\s_3 - \s_4)^2 + 2 a_{12} (\s_3 -
\s_4) (\s_3 + \s_4) \,,
\ee
which we write as
\be
\fbox{$ \displaystyle
ds^2_5  = h^2 d\q^2 + b^2 \left(\s_1^2 + \s_2^2 \right) + c^2 \left(f
\s_3 + \s_4 \right)^2 + g^2 \s_4^2 \,,$
}
\ee
with
\be
c^2 = \frac{(a_{22} - a_{11})^2}{a_{11}+a_{22}+2 a_{12}} \,, \qquad
g^2 = \frac{4(a_{22} a_{11} - a_{12}^2)}{a_{11}+a_{22}+2 a_{12}} \,,
\qquad f = \frac{a_{11} + a_{22} + 2 a_{12}}{a_{11} - a_{22}} \,.
\ee
The second form of the metric written here is useful for calculation
using the obvious vielbein
\be\label{eq:vielbeins}
e^1 = h d\q \,, \qquad e^2 = b \s_1 \,, \qquad e^3 = b \s_2 \,, \qquad
e^4 = c (f \s_3 + \s_4) \,, \qquad e^5 = g \s_4 \,.
\ee
The left-acting isometries follow from writing the metric in terms
of left-invariant forms. The remaining $U(1)$ symmetry is the
usual right-acting $U(1)$ isometry of the squashed three-sphere.
It is present because the metric does not depend on $\eta$.

\subsection{Moduli space of metrics: Volumes and Weyl eigenvalues}

Figure 2 shows the discrete moduli space of Einstein metrics
with $1 < \nu_1, \nu_2 < 100$. Each point corresponds to an
Einstein metric.
\begin{figure}[h]
\begin{center}
\vspace{4mm}
\epsfig{file=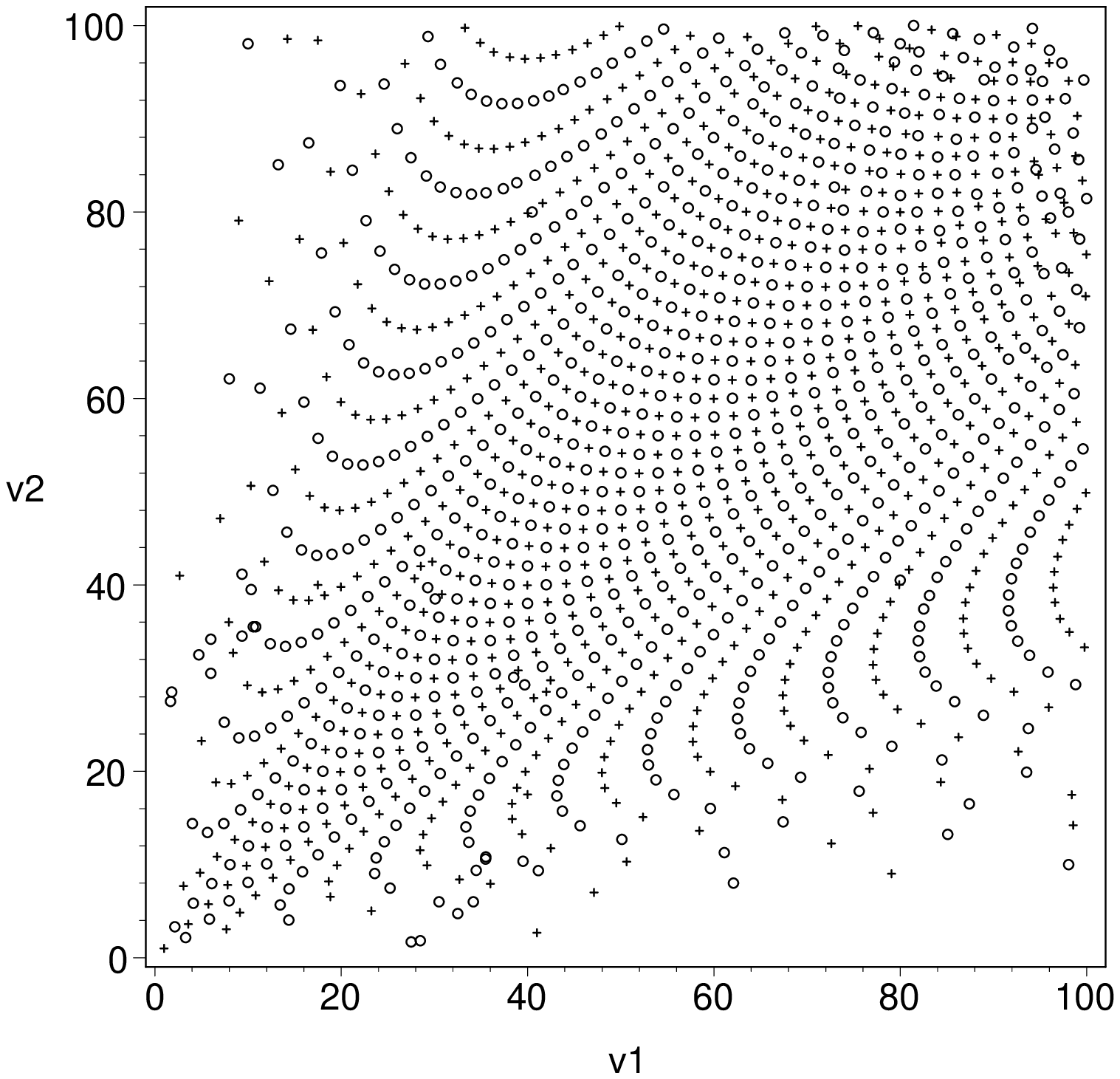,width=10cm}
\end{center}
\noindent {\bf Figure 2:} Moduli space of Einstein metrics.
Crosses correspond to topology $S^3\times S^2$. Circles have the
topology of the nontrivial $S^3$ bundle over $S^2$.
\end{figure}

So far we have parameterised the moduli space in terms of
$\nu_1,\nu_2$ or $k_1,k_2$. Although these quantities were useful for
discussing various limits, they do not have an invariant geometric
meaning. In this subsection we will consider how two geometric
properties of the metrics vary
as we move around the moduli space. The two properties will be the
volume of the manifold and the eigenvalues of the Weyl tensor. The
reason for choosing these quantities is that they are related to the
physics of generalised black holes and Freund-Rubin compactifications
constructed using these metrics. We will elaborate on physical
implications in the next subsection.

The volume of the manifolds is straightforward to calculate and is
given by
\be
\text{Vol}(\nu_1,\nu_2) = \frac{(\nu_1^2 \nu_2^2-1)^{5/2}
(\nu_1^2+\nu_2^2-2)^{3/2}\pi^3
}{(1+\nu_1^4\nu_2^2+\nu_1^2\nu_2^4-3\nu_1^2\nu_2^2) (\nu_2^2 +
\nu_1^2 \nu_2^2-2)(\nu_1^2 +
\nu_1^2 \nu_2^2-2)} \,.
\ee
The eigenvalues of the Weyl tensor acting on symmetric tracefree
tensors are given by
\be\label{eq:Weylaction}
C_a{}^c{}_b{}^d h_{c d} = \kappa h_{a b} \,.
\ee
For inhomogeneous manifolds the eigenvalues $\kappa=\kappa(x)$ depend
on the position. The maximum value taken by an eigenvalue on the
manifold, $\kappa_0$, gives a lower bound for the Lichnerowicz
spectrum\cite{Gibbons:2002th}. In five dimensions and with our
normalisation for the Einstein metrics, the bound is
\be\label{eq:Lich}
\Delta_L \geq 4 ( 5 - \kappa_0 ) \,.
\ee
The classical stability of generalised black holes and Freund-Rubin
compactifications depends on the Lichnerowcz spectrum, so $\kappa_0$
is an interesting quantity to consider.

Let us assume that $\nu_2 \geq \nu_1$. Results for the opposite case
may be obtained by interchanging $\nu_1$ and $\nu_2$. In the
present case, it turns out that the maximum value of the Weyl
eigenvalues is achieved at $\q=0$ and may be shown to be given by
a fairly simple expression
\be\label{eq:eval}
\kappa_0 =
\frac{(\nu_2^2-1)^2 (2+2\nu_1^2+\sqrt{9-2\nu_1^2+9\nu_1^4})}
{(\nu_1^2-1)(\nu_1^2 \nu_2^2-1)} \,.
\ee
To derive this expression, one should work with an orthonormal set
of vielbeins (\ref{eq:vielbeins}) and consider a basis of fourteen symmetric tracefree
matrices. The action of the Weyl tensor (\ref{eq:Weylaction}) on
this basis will then define a matrix whose eigenvalues are
required.

We can gain some intuition about instabilities by considering the
eigenmode corresponding to the
eigenvalue (\ref{eq:eval}). At $\q=0$, and using the tangent space
basis given by the vielbeins (\ref{eq:vielbeins}) we have
\be
h^0_{\bar{a}\bar{b}} = \left(
\begin{array}{ccccc}
 4 & & & & \\
 & -3 +  x & & & \\
 & &-3 + x  & & \\
 & & & -2 - 2 x& \\
 & & & & 4
\end{array}
 \right) \,,
\ee
where $x = 3 \nu_1^2 - \sqrt{9-2\nu_1^2+9\nu_1^4}$.
One may check that $C_a{}^c{}_b{}^d h^0_{c d} = \kappa_0 h^0_{a
 b}$ at $\q=0$. We expect that for the unstable spacetimes the
unstable mode will be concentrated near $\q=0$ and will be well
approximated by $h^0$ at that point. It is curious that the mode does
not depend on $\nu_2$.

Table 1 shows the values of the volume and the maximum Weyl
eigenvalue in two limits.

\begin{table}[h]
\begin{center}
  \begin{tabular}{|c||c|c|} \hline
\textsc{Limit} & \textsc{Volume} & \textsc{Weyl eigenvalue}  \\ \hline \hline
\rule[-5mm]{0mm}{13mm} Homogeneous: $\nu_1 = \nu_2 = \nu$ & $\displaystyle
\frac{(1+\nu^2)^{5/2} 2^{3/2}\pi^3}{(2+\nu^2)^2(2\nu^2+1)}$
&  $\displaystyle \frac{2+2\nu^2+\sqrt{9-2\nu^2+9\nu^4}}{1+\nu^2}$
\\ \hline

\rule[-5mm]{0mm}{13mm} `Rational limit': $\displaystyle \begin{array}{l} \nu_1, \nu_2 \to \infty \\
\nu_2/\nu_1 = q \geq 1 \text{ fixed} \end{array}$ & $\displaystyle \sqrt{1+\frac{1}{q^2}}
\frac{\pi^3}{\nu_1}$ & $5 q^2$  \\ \hline
  \end{tabular}

\vspace{0.3cm}
{\bf Table 1:} Volumes and maximum of Weyl eigenvalues in various limits.
\end{center}
\end{table}

We see that in the rational limits,
including the homogeneous limit $\nu \to \infty$, the volume goes
to zero. On the other hand, the Weyl
eigenvalue remains finite and may be tuned to any value allowed by
the integrality constraints: $k_i \in \Z^+$.

\subsection{Physical consequences: Entropy and stability}

Five dimensional compact Einstein manifolds appear in a simple way in
two contexts in higher dimensional gravity. Firstly, they can be
used to construct generalised black hole spacetimes in seven
dimensions
\be\label{eq:bh}
ds^2 = -f(r) dt^2 + \frac{dr^2}{f(r)} + r^2 ds^2_5 \,
\ee
with $f(r) = 1 - (1+r_+^2/L^2)(r_+/r)^4 + r^2/L^2$. This is a
solution to Einstein's equations, with a possible negative
cosmological constant $-1/L^2$. The horizon is at $r=r_+$. The
volume of the five dimensional manifold becomes the area of the
event horizon and is therefore proportional to the entropy of the
black holes.

The classical stability of generalised black holes has been
studied recently, both with vanishing cosmological constant
\cite{Gibbons:2002pq,Gibbons:2002th} and with a negative
cosmological constant \cite{Hartnoll:2003as}. The latter case has
interesting field theory implications using the AdS/CFT
correspondence \cite{Hartnoll:2004kz}.

For generalised black holes there is a simple criterion for classical
stability. The criterion depends on the minimum Lichnerowicz
eigenvalue of the horizon. For a five dimensional horizon, and
with vanishing cosmological constant, the criterion is
\cite{Gibbons:2002pq}
\be\label{eq:nocosmo}
\Delta_L \geq 4 \qquad \Leftrightarrow \qquad \text{stability}\,.
\ee
Therefore, from (\ref{eq:Lich}) we see that if any of the metrics
have $\kappa_0 \leq 4$, they will result in stable spacetimes.

For large generalised black holes with a negative cosmological
constant, $r_+/L >> 1$, the stability criterion is
\cite{Hartnoll:2003as}
\be\label{eq:ads}
\Delta_L \geq - A^2 \times \frac{r_+^2}{L^2} \qquad
\Leftrightarrow \qquad \text{stability} \,,
\ee
where $A^2$ is a positive ${\mathcal{O}}(1)$ coefficient that may be
determined numerically \cite{Hartnoll:2003as}.
In order to be unstable, the maximum Weyl eigenvalue $\kappa_0$
must therefore be very large.

A second type of solution that uses Einstein manifolds are
Freund-Rubin compactifications with a $5$ form field strength
\bea
ds^2 & = & ds^2_{\text{AdS}_{D-5}} + ds^2_5 \,, \nonumber \\
F_5 & = & \left[\frac{8(D-2)}{D-6} \right]^{1/2}
\text{vol}(M^5)
\,.
\eea
Remarkably the stability of these spacetimes
\cite{DeWolfe:2001nz,Gibbons:2002th} is according to precisely the
same criterion as for generalised black holes (\ref{eq:nocosmo}).
This may be due to the relation between generalised black holes and
Freund-Rubin compactifications pointed out in the introduction.

The relation between the Lichnerowicz spectrum and the eigenvalues
of the Weyl tensor (\ref{eq:Lich}) is such that it allows us to
prove the stability of the above spacetimes, but not instability.
However, it was found in \cite{Gibbons:2002th} that large positive
values of $\kappa_0$ tend to suggest unstable spacetimes.

One can check from (\ref{eq:eval}) that $\kappa_0 > 4$ for all
values except $\nu_1 = \nu_2 = \nu = 1$, which has precisely
$\kappa_0=4$. Therefore the only rigorous statement about
stability we can make is that the $\nu=1$ metric gives marginally
stable spacetimes. This metric is non other than $T^{1,1}$ so we
have reproduced a known fact \cite{Gubser:2001zr}. However, it
seems very likely that all the remaining metrics will give
unstable spacetimes. The instability means that these metrics have
limited interest for the AdS/CFT correspondence, but give rise to
interesting physics of generalised black holes.

Like the B\"ohm metrics \cite{Gibbons:2002th,bohm}, the moduli
space here includes regions where the maximum Weyl eigenvalue is
becoming arbitrarily large. For example, $q$ is arbitrary in
Table 1. This suggests that the lower bound on the Lichnerowicz
spectrum may become arbitrarily negative. This in turn implies
that the metrics give not only unstable flat space black holes,
but also unstable large black holes in anti-de Sitter space
according to the criterion (\ref{eq:ads}).

Unlike the B\"ohm metrics however, the metrics we consider here
are given explicitly, without need of numerical calculations.
Therefore it is substantially easier to study quantum fields on
these backgrounds. We begin this study below by computing the zeta
function of a free scalar field in some cases. Ultimately, one
would like to understand better the nature of the field theory
instability that seems to appear when the maximum Weyl eigenvalue
becomes large \cite{Hartnoll:2004kz}.

\section{Laplacian spectrum}

In this section we show how the Laplace eigenvalue equation for the
metrics we are considering separates. In the homogeneous cases we
find the spectrum explicitly. The equation we need to solve is
\be
-\Box \p = \lambda \p \,,
\ee
with
\be\label{eq:laplacian}
\Box = \Box_\q + \frac{1}{b(\q)^2 \sin\chi} \frac{\pa}{\pa \chi} \sin\chi \frac{\pa}{\pa
\chi} + \frac{1}{b(\q)^2} \left(e_3\right)^2 + a^{i j}(\q)
\frac{\pa}{\pa \psi^i} \frac{\pa}{\pa \psi^j}\,,
\ee
where
\be
e_3 = \frac{1}{\sin\chi} \left( \frac{\pa}{\pa \eta} - \cos\chi
\left[
\frac{\pa}{\pa \psi^1} + \frac{\pa}{\pa \psi^2} \right]
\right)\,,
\ee
and
\be
\Box_\q = \frac{1}{h(\q) b(\q)^2 \sqrt{\det a(\q)}} \frac{\pa}{\pa
\q} \left[ \frac{b(\q)^2 \sqrt{\det a(\q)}}{h(\q)} \frac{\pa}{\pa
\q} \right] \,,
\ee
where $a^{ij}$ is the inverse matrix of $a_{ij}$.

The eigenvalue equation can be separated by using the left acting
$SU(2)\times U(1)$ isometry to rewrite the Laplacian in terms of
the symmetry generators
\bea
\xi_1 & = & -\cot\chi \sin\left(\frac{\psi^1+\psi^2}{2}\right)
\left[ \frac{\pa}{\pa \psi^1} + \frac{\pa}{\pa \psi^2} \right] +
\cos\left(\frac{\psi^1+\psi^2}{2}\right) \frac{\pa}{\pa \chi} +
\frac{\sin\left(\frac{\psi^1+\psi^2}{2}\right)}{\sin\chi}
\frac{\pa}{\pa \eta} \,, \nonumber \\
\xi_2 & = & -\cot\chi \cos\left(\frac{\psi^1+\psi^2}{2}\right)
\left[ \frac{\pa}{\pa \psi^1} + \frac{\pa}{\pa \psi^2} \right] -
\sin\left(\frac{\psi^1+\psi^2}{2}\right) \frac{\pa}{\pa \chi} +
\frac{\cos\left(\frac{\psi^1+\psi^2}{2}\right)}{\sin\chi}
\frac{\pa}{\pa \eta} \,, \nonumber \\
\xi_3 & = & \frac{\pa}{\pa \psi^1} + \frac{\pa}{\pa \psi^2} \,,
\nonumber \\
\xi_4 & = & \frac{\pa}{\pa \psi^1} - \frac{\pa}{\pa \psi^2} \,.
\eea
Define the $SU(2)$ quadratic Casimir
\bea
\xi^2 & \equiv & \xi_1^2 + \xi_2^2 + \xi_3^2  \nonumber \\
 & = & (e_3)^2 + \frac{1}{\sin\chi} \frac{\pa}{\pa \chi} \sin\chi \frac{\pa}{\pa
\chi} + \left[\frac{\pa}{\pa \psi^1} + \frac{\pa}{\pa
\psi^2}\right]^2 \,.
\eea
The Laplacian may thus be written as
\be
\Box  =  \Box_\q + \frac{1}{b(\q)^2} \left[ \xi^2 - \xi_3^2 \right]
+ \frac{a^{11}}{4} \left[\xi_3 + \xi_4 \right]^2 +
\frac{a^{22}}{4} \left[\xi_3 - \xi_4 \right]^2 + 2\frac{a^{12}}{4}
\left[\xi_3 + \xi_4 \right] \left[\xi_3 - \xi_4\right] \,.
\ee
Now note that $\xi^2, \xi_3, \xi_4$ all commute with each other.
We know their eigenvalues from $SU(2)$ group theory. We must be
careful to account for the coordinate ranges correctly. Let $\p$
be a simultaneous eigenfunction. Then
\bea\label{eq:ranges}
\frac{1}{2} \left[\xi_3 + \xi_4 \right] \p & = -i n_1 \p \,, & \qquad\qquad\qquad n_1 \in \frac{k_1}{2} \Z \,,
\nonumber \\
\frac{1}{2} \left[\xi_3 - \xi_4 \right] \p & = -i n_2 \p \,, & \qquad\qquad\qquad n_2 \in \frac{k_2}{2} \Z \,,
\nonumber \\
\xi^2 \p & = -j(j + 1) \p \,, & \qquad\qquad\qquad j \in \frac{1}{2}
\Z^+\cup \{0\} \,.
\eea
The range of $n_1+n_2$ is restricted by $-j \leq n_1+n_2 \leq j$.
All the eigenvalues have a degeneracy of $2[j]+1$, where $[x]$ denotes
the integer part of $x$. The allowed values of
$n_1+n_2$ follow from the standard angular momentum
result $n_1+n_2 \in \{ -j, -j+1, -j+2, \ldots , j\}$ and then
retaining only
the values for $n_1+n_2$ that are consistent with
(\ref{eq:ranges}). We state the allowed values more explicitly below in
(\ref{eq:relate}) for the homogeneous case.

One is left with an ordinary differential equation for the $\q$
dependence of the eigenstate
\be\label{eq:thetaeqn}
\fbox{ $\displaystyle
- \Box_\q \p + \frac{j(j+1) - (n_1+n_2)^2}{b^2}\p + a^{ij}n_i
n_j\p = \lambda \p \,.$
}
\ee
Solutions of this equation will introduce another discrete parameter
$l$ labelling the eigenvalues. Write the eigenvalues of the full
Laplacian as $\lambda_{j,n_1,n_2,l}$. The zeta function for the
Laplacian on the Einstein manifolds is thus
\be
\zeta_{-\Box}(s) = \sum_j \sum_{n_1, n_2} \sum_l \frac{2[j] +
  1}{\lambda_{j,n_1,n_2,l}^s} \,,
\ee
with the summations restricted by (\ref{eq:ranges}) and the comments following.

\subsection{Homogeneous metrics: spectrum}

In the homogeneous case one has $\nu_1=\nu_2\equiv\nu$ and
\be
k_1=k_2=k=\nu \frac{\nu^2+2}{2\nu^2+1} \,.
\ee
The Laplacian simplifies. The relevant
terms are collected in appendix A. The equation becomes
\bea
-\frac{1}{\sin\q\cos\q} \frac{\pa}{\pa \q}
\left[\sin\q\cos\q \frac{\pa}{\pa\q}\right] \p +  4
\frac{(2\nu^2+1)^2}{(\nu^2+2)^2} \left[\frac{n_1^2}{\nu^2 \sin^2\q} +
\frac{n_2^2}{\nu^2 \cos^2\q} \right] \p =
\nonumber \\
- 4 \frac{(2\nu^2+1)}{\nu^2+2} j(j+1) \p +
6\frac{(2\nu^2+1)}{(\nu^2+2)^2} (n_1 + n_2)^2 \p +
\frac{(1+\nu^2)}{\nu^2+2}\lambda \p
\,.
\eea
This equation may be solved in terms of hypergeometric functions. If
one rewrites the equation as
\be
-\frac{1}{\sin\q\cos\q} \frac{\pa}{\pa \q}
\left[\sin\q\cos\q \frac{\pa}{\pa\q}\right] \p + \left[\frac{A^2}{\sin^2\q} +
\frac{B^2}{\cos^2\q} \right] \p
= \Lambda \p \,,
\ee
then two linearly independent solutions are
\be\label{eq:homosol}
\p_{\pm} = \cos^{\pm B}\q\sin^{A}\q {}_2F_1\left(
\frac{1+A\pm B - (1+\Lambda)^{1/2}}{2} , \frac{1+A\pm B +
(1+\Lambda)^{1/2}}{2} , 1 \pm B  ; \cos^2\q \right)\,.
\ee
Regularity at $\q=\pi/2$ requires that $\pm B = |B|$. Regularity
as $\q\to 0$ requires both $A = |A| \geq 0$ and further that the
hypergeometric function be polynomial. We will see below that the
solutions are in fact Jacobi polynomials. This condition for
(\ref{eq:homosol}) to be a polynomial is that
\be
\frac{1+|A|+|B| \pm (1+\Lambda)^{1/2}}{2} = - L \,, \qquad \qquad L \in
  \Z^+\cup \{0\}.
\ee
The solution of this equation is $\Lambda = 4 l (l+1) $
with
\be\label{eq:Lvals}
l = L + \frac{2\nu^2+1}{\nu(\nu^2+2)} ( |n_1| + |n_2| ) \,.
\ee
The eigenvalues of the Laplacian are thus seen to be
\be\label{eq:eigenvalues}
\fbox{ $\displaystyle
\lambda_{j,n_1,n_2,l} = \frac{4(2\nu^2+1)}{\nu^2+1} j(j+1) +
\frac{4(\nu^2+2)}{\nu^2+1} l(l+1) -
\frac{6(2\nu^2+1)}{(\nu^2+1)(\nu^2+2)} (n_1+n_2)^2 \,. $
}
\ee
In the following subsection we clarify the values that
$n_i,j,l$ may take. We will recover the spectrum
(\ref{eq:eigenvalues}) from a semiclassical quantisation of geodesic
energies in section 5.2 below.

\subsection{Towards the zeta function}

To calculate the zeta function later it will be convenient to parameterise
the eigenvalues with the following four integers
\be\label{eq:integers}
J, L \in \Z^+\cup \{0\} \,, \qquad \qquad  N_1, N_2 \in \Z \,.
\ee
The eigenvalues (\ref{eq:eigenvalues}) are then calculated using
\bea\label{eq:relate}
n_i & = & \frac{k}{2}N_i \,, \nonumber \\
j & = & \frac{k}{2} |N_1 + N_2| + J \,, \nonumber \\
l & = & \frac{1}{2} \left( |N_1| + |N_2| \right) + L \,.
\eea
The zeta function may be written
\be
\zeta_{-\Box}(s) = \sum_{L=0}^{\infty} \sum_{J=0}^{\infty}
\sum_{N_1=-\infty}^{\infty} \sum_{N_2=-\infty}^{\infty} \frac{2[k
    |N_1+N_2|/2]+2J+1}{\lambda_{N_1,N_2,J,L}^s}\,,
\ee
In the following section we will use this parameterisation to
calculate explicitly the zeta function for the homogeneous metrics
as $\nu \to \infty$. These are the metrics $T^{p,1}$ with $p$
large.

\subsection{Inhomogeneous metrics}

The differential equation for the $\q$ dependent part of the
solutions (\ref{eq:thetaeqn}) is significantly more complicated in
the general inhomogeneous case. The first step towards extracting
information from the equation is to clarify the structure of the
singular points of the equation.

Let us write $z=\cos 2\theta$. Then (\ref{eq:thetaeqn}) becomes
\beq\label{eq:form}
\frac{d^2}{dz^2}\phi+P(z)\frac{d}{dz}\phi+Q(z)\phi=\w(z)\lambda \phi \,,
\eeq
where
\beq\label{eq:P}
P(z)=\frac{1}{z-1}+\frac{1}{z+1}+\frac{1}{z-z_0} \,,
\eeq
and
\beq\label{eq:coeffs}
Q(z)-\w(z) \lambda =\frac{Q_1}{(z-1)^2}+\frac{Q_2}{(z+1)^2}+\frac{Q_3}{(z-z_0)^2}
+\frac{R_1}{z-1}+\frac{R_2}{z+1}+\frac{R_3}{z-z_0} \,,
\eeq
with $z_0=2\cos^2 \theta_0-1$ and
\beq
\cos^2 \theta_0=\frac{(1-\nu_2^2)(2-\nu_1^2-\nu_1^2 \nu_2^2)}
{(\nu_1^2-\nu_2^2)(1-\nu_1^2 \nu_2^2)} \,.
\eeq
The pole $z_0$ arose as a root of $\Delta_{\theta}=0$ in
\cite{Hashimoto:2004kc}. The coefficients $Q_i$ are given by
\beq\label{eq:Q}
Q_1=-\left(\frac{n_1}{k_1} \right)^2,
\quad Q_2=-\left(\frac{n_2}{k_2} \right)^2 \,,
\eeq
and
\beq\label{eq:Q1}
Q_3=-\frac{1}{(2-\nu_1^2-\nu_2^2)(1-\nu_1^2 \nu_2^2)}
\left( \frac{\nu_1(1-\nu_2^2)n_1}{k_1}+\frac{\nu_2(1-\nu_1^2)n_2}{k_2}
\right)^2 \,.
\eeq
The other coefficients $R_i$ satisfy a relation $R_1+R_2+R_3=0$
since $Q(z)=o(1/z^2)$. The explicit form of these coefficients is
somewhat complicated, so we leave them to appendix B.
Thus (\ref{eq:form}) is a Fuchs type
differential equation with 4 regular singular points including
$\infty$. Such equations have a canonical form studied by Heun. We
transform (\ref{eq:form}) to the canonical form in appendix E.

From (\ref{eq:P}) and (\ref{eq:Q}), the eigenfunction
around $z=1$ ($\theta=0$) takes the form
\beq
\phi(z)=(1-z)^{\mid n_1/k_1 \mid}f(z)
\eeq
with $f(z)$ an analytic function at $z=1$, while the eigenfunction
around $z=-1$ ($\theta=\pi/2$) takes the form
\beq
\phi(z)=(1+z)^{\mid n_2/k_2 \mid}g(z)
\eeq
with $g(z)$ an analytic function at $z=-1$. In the homogeneous case, we
saw that the regular function is given by a polynomial.

At infinity, the regular singular point implies that solution behaves
as
\be\label{eq:infty}
\phi(z)=z^{-1\pm\sqrt{1-Q_1-Q_2-Q_3+(z_0-1)R_1+(z_0+1)R_2}}h(z)
\ee
where $h(z)$ is analytic at infinity. For the homogeneous metrics $z_0
\to \infty$, and (\ref{eq:infty}) needs to be recalculated. However,
infinity remains a regular singular point in the homogeneous limit.

\subsection{Slightly inhomogeneous metrics: spectrum}

In this subsection, we compute the Laplacian spectrum for metrics that
are slightly inhomogeneous. We can do this using Quantum Mechanical
perturbation theory. In the rational limit at least, there are metrics
sufficiently near the homogeneous metrics in $\nu_i$ space for which
perturbation theory is applicable.

The perturbation we consider is
\be
\frac{\nu_2}{\nu_1}= 1+\epsilon \,, \qquad\text{with}\qquad \epsilon
\ll 1\,.
\ee
Equation (\ref{eq:form}) may be multiplied by $z^2-1$ so that it is
expressed in terms of a self-adjoint operator $H$. We will check
self-adjointness shortly.
\be\label{eq:Hdef}
H \phi \equiv (z^2-1) \left[ \frac{d^2}{dz^2} \phi(z)+
  P(z)\frac{d}{dz} \phi(z)+Q(z)\phi(z) \right]=\tilde{\w}(z)\lambda\phi(z) \,,
\ee
with $\tilde{\w}(z) = (z^2-1)\w(z)$. The coefficients are then perturbed to first order as
\be
P = P^{(0)} + \epsilon P^{(1)} \,, \qquad Q = Q^{(0)} + \epsilon
Q^{(1)} \,, \qquad \tilde{\w} = \tilde{\w}^{(0)} + \epsilon \tilde{\w}^{(1)} \,.
\ee
The zeroth order operator is thus
\beq
H^{(0)}=(z^2-1)\left( \frac{d^2}{dz^2}+ P^{(0)}(z) \frac{d}{dz}
+Q^{(0)}(z) \right) \,.
\eeq
Using the explicit form of the function
\be
P^{(0)}= \frac{1}{z-1}+\frac{1}{z+1},
\ee
the operator becomes
\beq
H^{(0)}=\frac{d}{dz}(z^2-1)\frac{d}{dz}+(z^2-1)Q^{(0)},
\eeq
which is manifestly self-adjoint with respect to the inner product
\beq
<\phi_1, \phi_2 >=\int_{-1}^{1}dz \phi_1(z) \phi_2(z).
\eeq
Expanding the eigenfunction
$\phi=\phi^{(0)}+\epsilon \phi^{(1)}$ and eigenvalue
$\lambda=\lambda^{(0)}+\epsilon \lambda^{(1)}$,
we have the well known result from first order perturbation theory
\be\label{eq:evalone}
\lambda^{(1)} = \frac{<\phi^{(0)}, (H^{(1)}- \lambda^{(0)} \tilde{\w}^{(1)})
  \phi^{(0)}>}{<\phi^{(0)},\tilde{\w}^{(0)} \phi^{(0)}>} \,.
\ee
In the present case we have
\beqa\label{eq:pertH}
H^{(1)} - \lambda^{(0)} \tilde{\w}^{(1)} &=& (z^2-1)\left(
P^{(1)}\frac{d}{dz} + Q^{(1)} - \lambda^{(0)} \omega^{(1)} \right)
\nonumber\\
&=& - \frac{\nu^2(1+\nu^2)}{(\nu^2-1)(\nu^2+2)}
(z^2-1)\frac{d}{dz} \nonumber \\
& + & Q_1^{(1)}\frac{z+1}{z-1}+Q_2^{(1)}\frac{z-1}{z+1}
+(R_1^{(1)}+R_2^{(1)})z+R_1^{(1)}-R_2^{(1)} \,,
\eeqa
the actual expressions for $R_i^{(1)}$ and $Q_i^{(1)}$ are still
rather large, so we relegate them to appendix B. The denominator
in (\ref{eq:evalone}) is given by
\be
\tilde{\w}^{(0)} = \frac{1+\nu^2}{4(2+\nu^2)} \,.
\ee

In calculating (\ref{eq:pertH}) one should keep $n_1,n_2,j,l$ fixed
under the perturbation $\nu_2/\nu_1= 1+\epsilon$. This is
because these are properties of the zeroth order solution, not of the full
equation. The relationship between $n_i,j,l$ and $N_i,J,L$ is kept to
be that of the homogeneous case (\ref{eq:relate}) and is not modified.

To evaluate (\ref{eq:evalone}), note that $\phi^{(0)}(z)$ is just the
homogeneous solution that we found previously (\ref{eq:homosol}). It
is convenient at this point to rewrite these solutions in terms of
Jacobi polynomials. It turns out that, up to an overall normalisation
\be
\phi^{(0)}(z) = (1-z)^{|N_1|/2} (1+z)^{|N_2|/2} P_L^{(|N_1|,|N_2|)}(z) \,.
\ee
We have used the integers introduced in (\ref{eq:integers}). We recall
that the Jacobi polynomials have the following definition
\beqa
P^{(\alpha,\beta)}_{n}(z)&=& \frac{(-1)^n}{2^n n!}(1-z)^{-\alpha}
(1+z)^{-\beta}
\left( \frac{d}{dz} \right)^{n}[(1-z)^{\alpha+n}(1+z)^{\beta+n}]
\nonumber\\
& \propto & F(-n, 1+n+\alpha+\beta,1+\alpha;(1-z)/2) \,.
\eeqa
The advantage of introducing Jacobi polynomials is that they satisfy
identities that will enable us to perform the various integrals
needed in the evaluation of (\ref{eq:evalone}). We see that we will
need to evalute $<\phi^{(0)},z
\phi^{(0)}>, <\phi^{(0)},1/(z-1) \phi^{(0)}>,<\phi^{(0)},1/(z+1)
\phi^{(0)}>$ and $<\phi^{(0)},(z^2-1)d/dz \phi^{(0)}>$. All of these
expressions may be computed, the required identities involving Jacobi
polynomials are given in appendix C.

The result is
\bea\label{eq:perturbed}
\lefteqn{<\phi^{(0)}, (H^{(1)}- \lambda^{(0)} \tilde{\w}^{(1)})
  \phi^{(0)}>} \nonumber \\
& & =  \frac{\nu^2(1+\nu^2)}{(\nu^2-1)(\nu^2+2)} \left[
    X_2 + \frac{|N_2|-|N_1|}{2} - X_1
    \left(L + \frac{|N_1|+|N_2|}{2} \right) \right] \nonumber \\
& & +  (R_1^{(1)} + R_2^{(1)}) X_1 +
  R_1^{(1)} - R_2^{(1)} +  Q_1^{(1)} + Q_2^{(1)} \nonumber \\
& & - \frac{|N_1| + |N_2| + 1 + 2L}{|N_1|}
  Q_1^{(1)} - \frac{|N_1| + |N_2| + 1 + 2L}{|N_2|} Q_2^{(1)} \,,
\eea
with the coefficients
\bea
X_1 & = & \frac{N_2^2-N_1^2}{(2L+|N_1|+|N_2|)(2L+|N_1|+|N_2|+2)} \,,
\nonumber \\
X_2 & = & \frac{L (|N_1|-|N_2|)}{2L+|N_1|+|N_2|} \,.
\eea
This expression does not appear to simplify to a pleasant expression in
terms of $n_i,j,l$. We will see in section 5.4 below how a very
similar result for the perturbed spectrum may be found from a
semiclassical quantisation of the classical geodesic energies.

\section{Zeta function for $T^{p,1}$ at large $p$}

We will consider the homogeneous metrics in the limit $\nu \to
\infty$, corresponding to $k\to\infty$ also.
The advantage of taking this limit is that summations of
discrete eigenvalues may be approximated as integrals over continuous
parameters. This fact will enable us to compute the zeta function.
There are infinitely many metrics in this limit, evenly
spaced in $\nu$. See Figure 2 above. Towards the end of this section
we will consider one physical application of the zeta function.

The spectrum becomes (\ref{eq:eigenvalues})
\be
\lambda = 8 j(j+1) + 4 l (l+1) - \frac{12}{\nu^2} (n_1 + n_2)^2 \,.
\ee
One should express the spectrum in term of the integers of (\ref{eq:integers})
\bea
n_i & = & \frac{\nu}{4} N_i \,, \nonumber \\
j & = & J + \frac{\nu}{4} | N_1 + N_2 | \,, \nonumber \\
l & = & L + \frac{1}{2} (|N_1| + |N_2|) \,.
\eea
where $J,L \in \Z^+\cup\{0\}$ and $N_i \in \Z$. The degeneracy of
the eigenvalues is $2[j]+1$.

In the large $\nu$ limit the spectrum may further be simplified to
\be\label{eq:largev}
\lambda = 8 \left[J + \frac{\nu}{4}|N_1 + N_2| \right]^2 + 4 \left[L +
\frac{1}{2} (|N_1| + |N_2|)
\right] \left[L +  \frac{1}{2} (|N_1| + |N_2|) + 1\right] \,.
\ee
This step does not hold when $N_1 + N_2 = 0$, we will consider
this case separately later. Until specified otherwise, assume that
$N_1 + N_2 \neq 0$. The degeneracy may now be taken to be
$2 J + \frac{\nu}{2}|N_1 + N_2|$.

The zeta function at large $\nu$ may therefore be written
\be
\zeta(s) = \sum_{N_1,N_2,J,L} \frac{ \nu^{1-2s} \left( 2
  \frac{J}{\nu}+\frac{1}{2}|N_1 + N_2|\right)} {\left(2 \left[2\frac{J}{\nu}
    + \frac{1}{2}|N_1 + N_2|
\right]^2 + \frac{4}{\nu^2} \left[L + \frac{1}{2} (|N_1| + |N_2|)
\right] \left[L +  \frac{1}{2} (|N_1| + |N_2|) + 1\right] \right)^s} \,.
\ee
Note that $\nu$ is large and $J$ only appears as $J/\nu$. We can
approximate this sum by an integral when $N_1 + N_2
\neq 0$. That is to say, we write
\be
\sum_{J=0}^{\infty} \to \nu \int_0^\infty d \left(\frac{J}{\nu}\right)
\,.
\ee
The integral is straightforward to perform and the result is
\be
\frac{-\nu^{2-2s}}{8(1-s)} \sum_{N_1,N_2,L} \left( \frac{1}{2}
  (N_1+N_2)^2 + \frac{4}{\nu^2} \left[L + \frac{1}{2} (|N_1| + |N_2|)
\right] \left[L +  \frac{1}{2} (|N_1| + |N_2|) + 1\right] \right)^{1-s} \,.
\ee

In order to do these sums, the modulus signs force us to consider
four cases separately. First note that we can take $N_1 \geq 0$
and the remaining eigenvalues have an extra degeneracy of 2, which
we will include at the end. The four cases then depend on the
value of $N_2$. The cases are: (I) $N_2 \geq 0$. (II) $N_2 =
-N_1$. (III) $-\infty < N_2 < -N_1$. (IV) $-N_1 < N_2 < 0$. The
total zeta function will be the sum of these contributions
\be\label{eq:thezeta}
\fbox{$\displaystyle
\zeta(s) = 2\left[\zeta_I(s) + \zeta_{II}(s) + \zeta_{III}(s) +
\zeta_{IV}(s)\right] \,. $
}
\ee
A check of our result is given in appendix D.

\subsection{Case I: $N_2 \geq 0$}

In this case we may write $M = N_1 + N_2 > 0$. We consider the
case $M=0$ later. The sum is expressed entirely in terms of M.
This introduces a degeneracy of $M+1$: the number of ways of
writing M as the sum of two positive integers. The zeta function
becomes
\be
\zeta_I(s) = \frac{\nu^{2-2s}}{8(s-1)} \sum_{M,L}
\frac{M+1}{\left(\frac{M^2}{2} + 4
  (\frac{L+M/2}{\nu})(\frac{L+M/2+1}{\nu})
  \right)^{s-1}} \,.
\ee
We may now convert the $L$ summation into an integral for the same
reasons as before. One obtains
\be
\zeta_I(s) = \frac{\nu^{3-2s}}{8(s-1)} \sum_{M=1}^{\infty}
\int_0^{\infty} \frac{M+1}{\left(\frac{M^2}{2} + 4
  (\frac{L+M/2}{\nu})(\frac{L+M/2+1}{\nu})
  \right)^{s-1}} d\left(\frac{L}{\nu}\right) \,.
\ee
This integral can be done using, for example, the Maple program.
The answer may be expressed in terms of hypergeometric functions.
Taking the large $\nu$ limit, one obtains
\be
\sum_{M=1}^{\infty} \frac{2^{s+1/2} \pi^{1/2}}{128} \nu^{3-2s}
\frac{\G(s-3/2)}{\G(s)} M^{3-2s} (1+M) \,.
\ee
But now the remaining sum is just a Riemann zeta function, so the
final answer for this case is
\be\label{eq:I}
\zeta_I(s) = \frac{2^{s+1/2} \pi^{1/2}}{128} \nu^{3-2s}
\frac{\G(s-3/2)}{\G(s)} \left[ \zeta_R(2s-3) + \zeta_R(2s-4) \right] \,.
\ee

\subsection{Case II: $N_2 = -N_1$}

In doing this case we will include the $N_1 = N_2 = 0$ case. In
these cases, the integral method is not valid. The zeta function
contribution from these values is
\be
\zeta_{II}(s) = \sum_{N_1,J,L} \frac{(2J+1)}{\left(8 J(J+1) + 4
  (L+N_1) (L+N_1 + 1) \right)^s} + {\mathcal{O}}(1/\nu^2) \,.
\ee
As usual, the summation should exclude the zero mode. This sum may
be simplified by considering $Q = L + N_1$. It may be possible to
do this sum using contour integration. This will not be necessary
here; the only point that is relevant is that to leading order
there is no $\nu$ dependence and that the first correction is
order $1/\nu^2$. This term will not give the dominant contribution
to physical quantities which will be at e.g. $s=0,-1/2$.

\subsection{Case III: $-\infty < N_2 < -N_1$}

Define the quantities $M=N_1+N_2 < 0$ and $N=N_1-N_2 > 0$. The
zeta function contribution for this case is
\be\label{eq:IIIa}
\zeta_{III}(s) = \frac{\nu^{2-2s}}{8(s-1)}\sum_{N,M,L}
\left[\frac{M^2}{2} + \frac{4}{\nu^2} (L+N/2)(L+N/2+1) \right]^{1-s}
\,.
\ee
What is the range of the summation for $M$ and $N$? If we fix $M$,
then we see that $N$ takes the values $N=|M|,|M|+2,|M|+4,\ldots$.
Therefore $L+N/2$ takes the values $|M|/2,|M|/2+1,|M|/2+2,\ldots$
with degeneracy $L+N/2+1-|M|/2$. Defining $x=(L+N/2)/\nu$ we may
rewrite the summation over $L$ and $N$ as an integral over $x$:
\be
\zeta_{III}(s) = \frac{\nu^{4-2s}}{8(s-1)} \sum_{M=-1}^{-\infty}
\int_{|M|/2v}^{\infty}
\frac{x+1/\nu-|M|/2\nu}{\left(\frac{M^2}{2}+4x(x+1/\nu)\right)^{s-1}} dx \,.
\ee
One may again do this integral. It is similar to the one
considered previously. The final result for the zeta function is
\be\label{eq:III}
\zeta_{III}(s) = \frac{- 2^s}{128}
\frac{\nu^{4-2s}\zeta_R(2s-4)\left[(s-1){}_2
  F_1(2,s,5-s;-1)+(s-4){}_2 F_1 (1,s-1,4-s;-1) \right]}{(s-1)(s-2)(s-3)(s-4)}  \,.
\ee
Note that this expression always has one power more in $\nu$ than
the contribution from case I (\ref{eq:I}). Therefore the
contribution from case I is negligible unless the expression
(\ref{eq:III}) vanishes. But if this expression were to vanish then
we would need to calculate the subleading term. This will not be a
problem in practice.

\subsection{Case IV: $-N_1 < N_2 < 0$ }

Define the quantities $M=N_1+N_2 > 0$ and $N=N_1-N_2 > 0$. The
zeta function contribution for this case is
\be
\zeta_{IV}(s) = \frac{\nu^{2-2s}}{8(s-1)}\sum_{N,M,L}
\left[\frac{M^2}{2} + \frac{4}{\nu^2} (L+N/2)(L+N/2+1) \right]^{1-s}
\,.
\ee
This is the same expression as before (\ref{eq:IIIa}) except that
now if we fix $M$ the range of $N$ is $M+2,M+4,...$. Now $L+N/2$
takes the values $M/2+1,M/2+2,...$, with degeneracy $L+N/2-M/2$.
Defining $x=(L+N/2)/\nu$ as before the zeta function becomes
\be
\zeta_{IV}(s) = \frac{\nu^{4-2s}}{8(s-1)} \sum_{M=1}^{\infty}
\int_{M/2\nu+1/\nu}^{\infty}
\frac{x-M/2\nu}{\left(\frac{M^2}{2}+4x(x+1/\nu)\right)^{s-1}} dx \,.
\ee
To leading order in $\nu$ this integral turns out to be the same
as in the previous case
\be
\zeta_{IV}(s) = \zeta_{III}(s) \,.
\ee

\subsection{Physical applications of the zeta function}

The most immediate application of the zeta function is to
calculate the thermodynamics of a free scalar field on the compact
manifold $M$. We can add a mass or coupling to the Ricci
scalar without changing the zeta function that we have calculated
to leading order in $\nu$, as can
be seen from (\ref{eq:largev}). We denote this coupling with $\kappa$.
More precisely, $\kappa$ will only be relevant when $|N_1+N_2|=0$, and
we saw that this case gives subleading contributions to the zeta function.

The free energy is given by the logarithm of the
partition function on $M \times S^1$, where the $S^1$
has length $\beta$ and the temperature of the scalar radiation is
$T=1/\beta$ as usual,
\bea
F & = & - \frac{1}{\b} \ln Z = - \frac{1}{\b} \ln \int {\mathcal{D}}\p\, e^{-
\int_0^{\beta} dt \int d^d x \sqrt{g}
  \left[\frac{1}{2} (\nabla \p)^2 + \frac{\kappa}{2} \p^2
  \right]} \nonumber \\
  & = & \frac{1}{2\b} \ln \det \frac{-(\pa/\pa t)^2-\square + \kappa}{\mu^2} \,,
\eea
where we have introduced an arbitrary mass scale $\mu$ so that the
dimensionalities are correct. Zeta function regularisation
\cite{Dowker:1975tf,Hawking:1976ja} then gives a finite expression
for the formal determinant
\be
F = - \frac{1}{2\b} \z_A'(0) \,,
\ee
where we have set $A = [-(\pa/\pa t)^2-\square + \kappa]/\mu^2$. The
zeta function is analytic at the origin, so the free energy is
finite.

We can calculate the free energy in the limits of low and high
temperature. In both cases, it is convenient to use the following
expression for the zeta function
\be\label{eq:zeta}
\z_A(s) = \Tr A^{-s} = \frac{1}{\G(s)} \int_0^{\infty} d\t \t^{s-1} \Tr e^{-\t
A} \,.
\ee
The eigenvalues of $(\pa/\pa t)^2$ on the $S^1$ factor are
\be
\w_n^2 = \left( \frac{2\pi}{\beta}\right)^2 n^2 \,, \quad n \in \Z
\,.
\ee
It follows that we may write the zeta function as
\be\label{eq:finiteT}
\z_A(s) = \frac{1}{\G(s)} \sum_{n=-\infty}^{\infty}
\int_0^{\infty}d\t \t^{s-1} e^{-\t \left(\frac{2\pi}{\mu\beta}\right)^2 n^2} \Tr_d e^{-\t A_S}
\,,
\ee
where $A_S$ refers to the operator acting on the $d$ dimensional
spatial section $M$.

First consider the theory at low temperature. In this regime we
may approximate the sum over the $S^1$ modes by an integral
\be
\sum_{n=-\infty}^{\infty} e^{-\t \left(\frac{2\pi}{\mu\beta}\right)^2
n^2} \to \frac{\b\mu}{2\pi} \int_{-\infty}^{\infty} dk e^{-\t k^2}
= \frac{\b\mu}{(4\pi)^{1/2}} \frac{1}{\t^{1/2}} \,.
\ee
The zeta function may thus be written as
\bea
\z_A(s) |_{\text{Low T}} & = & \frac{\b\mu}{\G(s)(4\pi)^{1/2}}
\int_0^{\infty}d\t \t^{s-1/2-1} \Tr_d e^{-\t A_S} \,, \nonumber \\
 & = & \frac{\b\mu}{(4\pi)^{1/2}} \frac{\G(s-1/2)}{\G(s)}
\z_{A_S}(s-1/2) \,.
\eea

The technique for calculating the high temperature behaviour was
introduced by \cite{Dowker:md}. For the terms in the sum with $n
\neq 0$, it turns out that we can use the Schwinger-de Witt
expansion for the heat kernel $\Tr_d e^{-\t A_S}$ as $\t \to 0$.
This will work when the temperature is larger than the curvatures,
because in this case the dominant contribution to the integral
comes from small values of $\t$. The Schwinger-de Witt asymptotic
expansion is
\be\label{eq:expansion}
\Tr_d e^{-\t A_S} \sim \frac{\mu^d}{(4\pi)^{d/2}} \sum_{k=0}^{\infty} \frac{a_{2k}}{\mu^{2k}} \t^{k-d/2}
= \frac{\mu^d}{(4\pi)^{d/2}} \left[ a_0 \t^{-d/2} + \frac{a_2}{\mu^2}
\t^{1-d/2} + \cdots \right] \,,
\ee
where the first few $a_{2k}$ are given in appendix D. If we let
${\frak{R}}^k$ denote a generic curvature scalar with mass
dimension $2k$, then we have that $a_{2k} =
{\mathcal{O}}({\frak{R}}^k)$.

Substituting the expansion (\ref{eq:expansion}) into the
expression for the zeta function (\ref{eq:finiteT}) and doing the
integral over $\t$, one obtains
\bea
\lefteqn{\z_A(s) |_{\text{High T}} \,  = \,  \z_{A_S}(s) +
  \frac{2\mu^d}{(4\pi)^{d/2} \G(s)} \sum_{n=1}^{\infty}
\sum_{k=0}^{\infty} \left(\frac{4\pi^2 n^2}{\mu^2 \b^2}
\right)^{d/2-s-k} \G(s+k - d/2) \frac{a_{2k}}{\mu^{2k}} \nonumber} \\
 & & =  \z_{A_S}(s) + \frac{2\mu^d}{(4\pi)^{d/2} \G(s)}
\sum_{k=0}^{\infty} \left(\frac{4\pi^2}{\mu^2 \b^2}
\right)^{d/2-s-k} \G(s+k - d/2) \z_R(2s+2k-d) \frac{a_{2k}}{\mu^{2k}} \,,
\eea
where $\z_R$ is the Riemann zeta function. We see that the series
expansion will be valid if $\mid a_{2k} \b^{2k}\mid
\ll 1$, that is, if the temperature is large compared with
curvature scales $\mid {\frak{R}}\mid \ll T^2$.

Thermodynamic quantities may be calculated using standard formulae
such as $E = \pa (\b F)  / \pa \b$ and $S = - \pa F / \pa T$. The
results in the low and high temperature limits are shown in Table
2. In this table we restore the dependence on the scalar curvature
$R$ for completeness. At low temperature, the leading contribution
comes from $\zeta_{III}(s)$ and $\zeta_{IV}(s)$.

\begin{table}[h]
\begin{center}
  \begin{tabular}{|c||c|c|} \hline
\textsc{Quantity} & \textsc{Low Temperature} & \textsc{High Temperature}  \\ \hline \hline
\rule[-5mm]{0mm}{13mm} $f$ & $\displaystyle - \frac{\sqrt{10 R}
  \nu^5}{2419200} + \cdots $
&  $\displaystyle - \frac{2 \sqrt{2} \pi^6}{945 \nu} T^6 + \frac{(6
  \kappa + R) \sqrt{2} \pi^4}{2160 \nu} T^4  + \cdots $
\\ \hline

\rule[-5mm]{0mm}{13mm} $E$ & $\displaystyle - \frac{\sqrt{10 R}\,
  \nu^5}{2419200} + \cdots $
&  $\displaystyle \frac{2\sqrt{2} \pi^6}{189 \nu} T^6 -
  \frac{(6\kappa+R) \sqrt{2} \pi^4}{720 \nu} T^4 \cdots $
\\ \hline

\rule[-5mm]{0mm}{13mm} $S$ & $\displaystyle 0 + \cdots $
&  $\displaystyle \frac{4\sqrt{2} \pi^6}{315 \nu} T^5 -
\frac{(6\kappa + R)\sqrt{2}\pi^4}{540 \nu} T^3 \cdots $
\\ \hline

  \end{tabular}

\vspace{0.3cm}
{\bf Table 2:} Thermodynamics of a scalar field on
$M\times S^1$ at high and low temperature.
\end{center}
\end{table}

\section{Hamilton-Jacobi equation and geodesics}

\subsection{Separability of the Hamilton-Jacobi equation}

In this section we show that the Hamilton-Jacobi equation for the
geodesics of the metrics we are considering is integrable.
Hence we obtain first order equations for the geodesics.

The Lagrangian for a free particle is
\be
{\mathcal{L}} = g_{ab} dx^a dx^b = h^2 \dot{\q}^2 + b^2 (\dot{\chi}^2
+ \sin^2\chi \dot{\eta}^2) + a_{ij} (\dot{\psi^i}+\cos\chi
\dot{\eta})(\dot{\psi^j}+\cos\chi \dot{\eta}) \,,
\ee
where a dot denotes differentiation with respect to some time
parameterisation of the geodesics.
The geodesic equations may be separated by using a
Hamilton-Jacobi method. In this description, the dual momenta are
\be
p_a = \frac{\pa{\mathcal{L}}}{\pa \dot{x^a}} = \frac{\pa S}{\pa x^a} \,.
\ee
The coordinates $\psi^i$ and $\eta$ are cyclic, giving three conserved
momenta, $p_{\psi^i} = J_i$ and $p_{\eta} = J_3$.
The first order equations for these coordinates are easily seen to be
\bea
\dot{\eta} & = & \frac{1}{2} \frac{(J_1+J_2) \cos\chi - J_3}{b^2
  \sin^2\chi} \,, \nonumber \\
\dot{\psi^i} & = & \frac{1}{2} a^{i j} J_j - \cos\chi \dot{\eta} \,.
\eea
To find the remaining geodesic equations, consider the Hamilton-Jacobi equation
\be
\frac{1}{h(\q)^2} \left(\frac{\pa S}{\pa \q}\right)^2 + \frac{1}{b(\q)^2}\left( \frac{\pa S}{\pa \chi}
\right)^2 + \frac{1}{b(\q)^2} \left(e_3 S \right)^2 + a^{i j}
\frac{\pa S}{\pa \psi^i} \frac{\pa S}{\pa \psi^j} = E \,,
\ee
with $E$ a constant. This equation may be separated by taking
\be\label{eq:separation}
S = W(\q) + U(\chi) + J_3 \eta + J_1 \psi^1 + J_2 \psi^2 \,.
\ee
One then obtains the equations
\be\label{eq:Udot}
\left(\frac{d U}{d\chi} \right)^2 + \frac{1}{\sin^2\chi} \left(J_3 - \cos\chi [J_1 + J_2]
\right)^2= L^2 \,,
\ee
and
\be\label{eq:Wdot}
\frac{1}{h^2} \left(\frac{d W}{d\q} \right)^2 + a^{ij}J_i J_j +
\frac{L^2}{b^2} = E \,,
\ee
where $L$ is a constant. Now use the relations
\bea
\frac{d U}{d\chi} & = & \frac{\pa S}{\pa \chi} =
\frac{\pa{\mathcal{L}}}{\pa \dot{\chi}} = 2 b^2 \dot{\chi} \,,
\nonumber \\
\frac{d W}{d\q} & = & \frac{\pa S}{\pa \q} =
\frac{\pa{\mathcal{L}}}{\pa \dot{\q}} = 2 h^2 \dot{\q} \,,
\eea
to obtain equations for $\dot{\chi}$ and $\dot{\q}$ from
(\ref{eq:Udot}) and (\ref{eq:Wdot}) respectively. These equations can
then be used to eliminate the time dependence and hence obtain
equations for the orbits. For example
\be
\frac{h^2}{b^2} \left(\frac{d\q}{d\chi}\right)^2 = \frac{E^2-L^2/b^2 - a^{ij}J_i
  J_j}{L^2\sin^2\chi - [J_3 - \cos\chi(J_1+J_2)]^2} \sin^2\chi \,.
\ee
Similar equations may be derived for $\psi^i$ and $\eta$ in terms of
$\chi$. In fact, one obtains a simple equation for $d\chi/d\eta$
\be\label{eq:magnetic}
\left(\frac{d\chi}{d\eta}\right)^2 = \sin^2\chi \left(
\frac{L^2\sin^2\chi}{(J_3-\cos\chi (J_1+J_2))^2}-1 \right) \,.
\ee
This equation describes the projection of the geodesic onto the
$(\chi,\eta)$ two-sphere. One can recognise
(\ref{eq:magnetic}) as describing the motion of a charged particle
on a sphere with a magnetic monopole background. This is
not surprising given that the full metric was constructed from a
principle $T^2$ bundle over the $S^2$. It is well known that
charged particle motion on a sphere with a magnetic monopole
background results in closed circular orbits.
A little algebra shows that the motion (\ref{eq:magnetic}) describes a
circle on the two sphere about an axis ${\bf n}$ with opening angle
$\Theta$. One finds
\be
\sin^2\Theta = \frac{L^2}{(J_1+J_2)^2+L^2} \,,
\ee
and
\be
{\bf n} = \sqrt{1 - \frac{J_3^2}{L^2}\sin^2\Theta }
\left[\sin\Xi \, {\bf e}_x + \cos\Xi\, {\bf e}_y \right] + \frac{J_3}{L} \sin\Theta
     {\bf e}_z \,,
\ee
where $\Xi$ is an arbitrary angle and $\{ {\bf e}_x,{\bf e}_y,{\bf
  e}_z\}$ is the standard Cartesian basis.

\subsection{Action-angle analysis of the homogeneous case:
semiclassical spectrum}

In this subsection we carry out a full action-angle variable analysis
of the geodesics of the homogeneous metrics. This analysis allows us to
compute the frequencies of closed geodesics and also allows a
semiclassical quantisation of the system. We see that the
semicalssical spectrum essentially agrees with the spectrum of the Laplacian
that we calculated previously.

We work in phase space
\be
M = T^{*}(S^2 \times S^3) \,,
\ee
with coordinates $(x^a)=(\b=2 \theta,\chi,\eta, \psi_1, \psi_2)$
and their dual momenta $p_a$. The canonical one form $\a$ and corresponding
symplectic form $\w$ are as always
\be
\alpha = p_a \wedge dx^a \,, \qquad \w = d\a \,.
\ee
The Hamiltonian for the system is
\be
H(x,p) = g^{ab}p_a p_b \,.
\ee
We may write $H$ explicitly as
\beqa\label{eq:hamil}
H&=&\frac{4(2+\nu^2)}{1+\nu^2}
\left( p_\b^2+\frac{1}{k^2 \sin^2\b}
((p_{\psi_2}+p_{\psi_1})^2
-2(p_{\psi_2}+p_{\psi_1})(p_{\psi_2}-p_{\psi_1})\cos\b
+(p_{\psi_2}-p_{\psi_1})^2) \right)\nonumber \\
&+&\frac{4(1+2\nu^2)}{1+\nu^2}
\left( p_\chi^2+\frac{1}{ \sin^2\chi}
(p_{\eta}^2
-2(p_{\psi_2}+p_{\psi_1})p_\eta \cos\chi
+(p_{\psi_2}+p_{\psi_1})^2) \right)\nonumber \\
&-&\frac{6(1+2\nu^2)}{(1+\nu^2)(2+\nu^2)}(p_{\psi_2}+p_{\psi_1})^2 \,.
\eeqa
This Hamiltonian system is integrable since
there exist $5=\text{dim} M/2$ independent functions $F_a$ $(a=1\ldots
5)$ such that
\begin{itemize}
\item $\{F_a, H \}_{\omega} =0$,

\item $\{F_a, F_b \}_{\omega} =0$.
\end{itemize}
The functions may be taken to be
\beqa\label{eq:constants}
F_1^2&=&p_\b^2+\frac{1}{k^2 \sin^2\b}
((p_{\psi_2}+p_{\psi_1})^2
-2(p_{\psi_2}+p_{\psi_1})(p_{\psi_2}-p_{\psi_1})\cos\b
+(p_{\psi_2}-p_{\psi_1})^2)\nonumber\\
F_2^2&=&p_\chi^2+\frac{1}{ \sin^2\chi}
(p_{\eta}^2
-2(p_{\psi_2}+p_{\psi_1})p_\eta \cos\chi
+(p_{\psi_2}+p_{\psi_1})^2)\nonumber\\
F_3&=& p_{\eta},\qquad  F_4= p_{\psi_1}, \qquad F_5= p_{\psi_2}.
\eeqa
In which case the Hamiltonian (\ref{eq:hamil}) may be written
\be
H = \frac{4(2+\nu^2)}{1+\nu^2} F_1^2 + \frac{4(1+2\nu^2)}{1+\nu^2} F_2^2
-\frac{6(1+2\nu^2)}{(1+\nu^2)(2+\nu^2)}(F_4+F_5)^2 \,.
\ee
In fact, the quadratic conserved quantities $F_1$ and $F_2$ are related to
two reducible Staeckel-Killing tensors, $K_i^{a b}$, of the background
\be
F_i = K_i^{a b} p_a p_b  \qquad \qquad (i=1,2) \,.
\ee
This follows from writing the quantities as
\beq
F_1^2=\tilde{\xi}_1^2 +\tilde{\xi}_2^2 +\tilde{\xi}_3^2 \,,\quad
F_2^2=\xi_1^2 +\xi_2^2 +\xi_3^2 \,,
\eeq
where we have used the
generators of the $SU(2) \times SU(2)$ symmetry,
\beqa
\xi_1 &=& -\cot \chi \sin ((\psi_1+\psi_2)/2) (p_{\psi_1}+p_{\psi_2})
+\cos((\psi_1+\psi_2)/2) p_{\chi}+
\frac{\sin ((\psi_1+\psi_2)/2) }{\sin \chi} p_{\eta}\nonumber \,, \\
\xi_2 &=& -\cot \chi \cos ((\psi_1+\psi_2)/2) (p_{\psi_1}+p_{\psi_2})
-\sin((\psi_1+\psi_2)/2) p_{\chi}+
\frac{\cos ((\psi_1+\psi_2)/2) }{\sin \chi} p_{\eta}\nonumber \,, \\
\xi_3 &=& p_{\psi_1}+p_{\psi_2} \,,
\eeqa
and
\beqa
\tilde{\xi}_1 &=&
-\cot \b \sin (k(\psi_1+\psi_2)/2) (p_{\psi_1}+p_{\psi_2})/k
+\cos(k(\psi_1+\psi_2)/2) p_{\b}\nonumber\\
&-&\frac{\sin (k(\psi_1+\psi_2)/2) }{\sin \b}
(p_{\psi_1}-p_{\psi_2})/k \,,
 \nonumber\\
\tilde{\xi}_2 &=& -\cot \alpha \cos (k(\psi_1+\psi_2)/2)
(p_{\psi_1}+p_{\psi_2})/k
-\sin(k(\psi_1+\psi_2)/2) p_{\b} \nonumber\\
&-&\frac{\cos (k(\psi_1+\psi_2)/2) }{\sin \b}
(p_{\psi_1}-p_{\psi_2})/k \,,
\nonumber\\
\tilde{\xi}_3 &=& (p_{\psi_1}+p_{\psi_2})/k \,,
\eeqa
which satisfy the relations $\{\xi_i,\xi_j\}=-\epsilon_{ijk}\xi_k$ and
$\{\tilde{\xi}_i,\tilde{\xi}_j\}=-\epsilon_{ijk}\tilde{\xi}_k$.

The five constants of motion allow us to consider the level set
\beq
M_f=\{(x,p): F_a(x,p)=f_a \}.
\eeq
General theorems show that $M_f$ is diffeomorphic to the 5 dimensional
torus. The action variables are constructed by considering 5
independent cycles in the torus, $C_a$, and writing
\be
I_a = \frac{1}{2\pi} \oint_{C_a} \alpha \,.
\ee
The definition is in fact independent of the cycle chosen to represent a
given homology class because the symplectic form $\omega = d\a$
vanishes when restricted to $M_f$.
From their definition, the $\{I_a\}$ variables will be invertible functions of
the constants $\{f_a\}$ only. If we invert this relationship, we may
consider $f_a(I)$. The key step in action-angle analysis is then to note
that the Hamilton-Jacobi function may be considered as a function of
$x^a$ and $I_a$: $S=S(x,I)$. Thus one may define new coordinates
\be
\phi^a=\frac{\partial S}{\partial I_a} \,.
\ee
These are the angle variables. The action-angle variables satisfy three
important properties. Firstly, $(\phi^a,I_a)$ are canonical coordinates on the
phase space
\be
\omega = d\phi^a \wedge dI_a \,.
\ee
Secondly, the variables $\phi^a$ are indeed angles on the cycles $C_a$
\be
\oint_{C_a} d\phi^b = 2\pi \delta^b_a \,.
\ee
Thirdly, the time evolution equations for $(\phi^a,I_a)$ are trivial
\bea\label{eq:trivial}
\dot{I_a} & = & 0 \,, \nonumber \\
\dot{\phi^a} & = & \frac{\pa H(I)}{\pa I_a} \equiv \Omega^a(I) \,
\Rightarrow \, \phi^a = \Omega^a(I) t + \text{const.}
\eea

The next step is to choose 5 cycles in $M_f$. Three cases are
particularly straightforward. Consider cycles that have tangent
vectors $\frac{\pa}{\pa \eta}$, $\frac{\pa}{\pa \psi_1}$ and
$\frac{\pa}{\pa \psi_2}$. The action variables are respectively,
using (\ref{eq:constants}),
\beq
\fbox{$\displaystyle
I_3=f_3, \quad I_i=\frac{2}{k}f_i \quad (i=4,5).$
}
\eeq
The remaining two integrals are slightly more complicated. They
are most easily computed by re-expressing the action variable as an
integral over a surface in phase space with $C_a =\pa S_a$
\be\label{eq:surface}
2\pi I_a = \oint_{C_a} \a = \int \int_{S_a} d\a = \int \int_{S_a} dp_b
\wedge dx^b  \,.
\ee
First we describe the curves. The two cycles will be taken to be
the intersection of $M_f$ with the surface generated by tangent
vectors $(\frac{\pa}{\pa \b}, \frac{\pa}{\pa p_\b})$ and
$(\frac{\pa}{\pa \chi}, \frac{\pa}{\pa p_\chi})$ respectively. From
(\ref{eq:constants}) we can calculate the equation for the cycles,
in both cases the curve takes the general form
\beq\label{eq:cycle}
C=\{(x,y)\in [0,\pi] \times \R \, | \, y^2=a^2-(b^2-2bc\cos
x+c^2)/\sin^2x \} \,.
\eeq
If the following two conditions hold
\beq\label{eq:conditions}
a^2-|b c| > 0, \qquad (a^2-b^2)(a^2-c^2)> 0 \,,
\eeq
then the equation for the curve becomes
\beq
y^2=a^2(\cos x - \delta_1)(\delta_2-\cos x)/\sin^2 x, \quad -1 \le
\delta_1 < \delta_2 \le 1 \,,
\eeq
with
\beq\label{eq:delta}
\delta_{1,2}=\frac{1}{a^2}(b c \pm \sqrt{(a^2-b^2)(a^2-c^2)}) \,.
\eeq
The curve $C$ is closed and the area $A$ enclosed by $C$ is
\beqa\label{eq:area}
A & = & 2|a|\int_{\delta_1}^{\delta_2}
\frac{\sqrt{(t-\delta_1)(\delta_2-t)}}{1-t^2}dt\nonumber\\
 & = & |a| \pi \left(2-\sqrt{(1+\delta_1)(1+\delta_2)}
-\sqrt{(1-\delta_1)(1-\delta_2)} \right)\nonumber\\
 & = & \pi(2 |a|-|b+c|-|b-c|)\,.
\eeqa
By (\ref{eq:surface}) we see that the action variables will be
given by the area inside the curve in the $(x,p)$ plane. Now
consider the two cases:

{\bf (A) $\, I_1(f)$:} In the equation for the curve (\ref{eq:cycle}) we have $a=f_1,
b=(f_4+f_5)/k, c=(f_5-f_4)/k$. We can take $f_1>0$ without loss of
generality. Then, $|I_1(f)|=A_1/2\pi$, where $A_1$ is the area
given by (\ref{eq:area})
\beq\label{eq:I1}
\fbox{$\displaystyle
|I_1|=f_1-\Bigl|\frac{f_4}{k} \Bigr|-\Bigl|\frac{f_5}{k} \Bigr|\,.$
}
\eeq

{\bf (B) $\, I_2(f)$:} Here we have $a=f_2, b=f_4+f_5, c=f_3$, and taken $f_2>0$ without
loss of generality. Then, $|I_2(f)| =A_2/2\pi$, where $A_2$ is the
area given by (\ref{eq:area}):
\beq\label{eq:I2}
\fbox{$\displaystyle
|I_2|=f_2-\frac{1}{2}(|f_4+f_5+f_3|+|f_4+f_5-f_3|)\,.$
}
\eeq
Here, we have assumed the condition (\ref{eq:conditions}) for
$f_a$, which is equivalent to that of the existence of periodic
orbits in $M$.

We can now calculate the frequencies of the orbits. The frequency
$\Omega^a$ is (\ref{eq:trivial})
\beq
\Omega^a=\frac{\partial H(I)}{\partial I_a}\,,
\eeq
where
\beq\label{eq:actionhamil}
H(I)=\frac{4(2+\nu^2)}{1+\nu^2}f_1^2(I)
+\frac{4(1+2\nu^2)}{1+\nu^2}f_2^2(I)
-\frac{3\nu^2(\nu^2+2)}{2(1+\nu^2)(1+2\nu^2)}(I_4+I_5)^2.
\eeq
Using (\ref{eq:I1})and (\ref{eq:I2}) we have
\beqa
f_1 &=& |I_1|+\frac{1}{2}(|I_4|+|I_5|)\,,\nonumber\\
f_2 &=& |I_2|+\frac{1}{4}(|k(I_4+I_5)+2 I_3|+|k(I_4+I_5)-2
I_3|)\,.
\eeqa
This is a classical version of the Laplace spectrum
(\ref{eq:eigenvalues}) written with the action coordinates.

We may use the expression for the Hamiltonian in terms of the action
variables (\ref{eq:actionhamil}) to perform a semiclassical
quantisation of the spectrum. The semiclassical prescription is to
replace action variables by integers. Consider the following
quantisation
\bea
I_1 \to L \,, \qquad I_3 \to M \,, \qquad
I_4 \to N_1 \,, \qquad I_5 \to N_2 \,, \nonumber \\
I_2 \to J + \frac{k}{2}|N_1 + N_2| - \frac{1}{4}(|k(N_1+N_2)+2 M|+|k(N_1+N_2)-2
M|)\,.
\eea
The slightly awkward quantisation of $I_2$ is in order to make contact
with our previous notation. It is not difficult to see that the above
expression sets $I_2$ to be an integer and $J$ to be a positive integer.

Using the definitions of (\ref{eq:relate}) we see that the
semiclassical spectrum may be written as
\be
H = \frac{4(2\nu^2+1)}{\nu^2+1} j^2 +
\frac{4(\nu^2+2)}{\nu^2+1} l^2 -
\frac{6(2\nu^2+1)}{(\nu^2+1)(\nu^2+2)} (n_1+n_2)^2 \,.
\ee
This is precisely the same as the spectrum for the Laplacian that we
calculated previously in (\ref{eq:eigenvalues}), up to the usual
semiclassical error $j^2 \to j(j+1)$, $l^2 \to l(l+1)$. This agreement
suggests an approach for calculating the Laplacian spectrum in the
inhomogeneous cases. We can understand the semiclassical disagreement
by comparing a naive quantisation of the classical Hamiltonian
(\ref{eq:hamil}), $p_a \to i\pa_{x^a}$, with the Laplacian
(\ref{eq:laplacian}). The quantised Hamiltonian has terms $\pa_{\b}^2$
and $\pa_{\chi}^2$ whilst the Laplacian has $1/\sin\b
\pa_{\b}[\sin\b \pa_{\b}]$ and $1/\sin\chi \pa_{\chi} [\sin\chi
\pa_{\chi}]$. This represents an ordering ambiguity in quantisation
and explains $j^2$ versus $j(j+1)$.

Finally, we can write down an explicit solution for the functions
$W(\b,I)$ and $U(\chi,I)$ that appeared in the separation of
the Hamilton-Jacobi equation (\ref{eq:separation}).
Using the following formula, which is the integral of
(\ref{eq:area}) in an indefinite form,
\beqa
K(t;\delta_1,\delta_2) & \equiv &
\int \frac{\sqrt{(t-\delta_1)(\delta_2-t)}}{1-t^2}dt \nonumber\\
&=& \sqrt{(1+\delta_1)(1+\delta_2)}
\arctan \left(
\sqrt{ \frac{(1+\delta_1)(\delta_2-t)}{(1+\delta_2)(t-\delta_1)}}
\right)\nonumber\\
&+&\sqrt{(1-\delta_1)(1-\delta_2)}
\arctan \left(
\sqrt{ \frac{(1-\delta_1)(\delta_2-t)}{(1-\delta_2)(t-\delta_1)}}
\right)\nonumber\\
&-& 2 \arctan
\left(
\sqrt{ \frac{\delta_2-t}{t-\delta_1}}
\right) \quad(-1 \le \delta_1 < \delta_2 \le 1),\nonumber
\eeqa
we find that
\beqa
W(\b,I)=f_1(I) K(\cos \b ; \delta_1(I),\delta_2(I))
,\nonumber\\
U(\chi,I)=f_2(I)
 K(\cos \chi ; \tilde{\delta}_1(I),\tilde{\delta}_2(I)),
\eeqa
where the parameters $\delta_i(I)$ and $\tilde{\delta}_j(I)$ are
determined as a function of $I_a$ by (\ref{eq:delta}) with the
values for $a,b,c$ given in (A) and (B) above, respectively. $W$
and $U$ are multivalued functions which have multiplicities $N
A_1$, $N A_2$, with $ N \in \Z$.

\subsection{Action-angle analysis of the inhomogeneous case}

After setting $z=\cos 2 \theta$, hence $p_{\theta}=-2 p_z \sqrt{1-z^2}$,
the Hamiltonian is given by
\beqa\label{eq:inhamil}
H &=& \frac{4(1-z^2)}{h^2(z)}p^2_z+\frac{1}{b^2(z)} \left(
p^2_{\chi}+\frac{1}{\sin^2 \chi}
(p^2_{\eta}-2(p_{\psi_2}+p_{\psi_1})p_{\eta} \cos \chi+
(p_{\psi_2}+p_{\psi_1})^2 \cos^2 \chi) \right) \nonumber\\
&+& a^{ij}(z)p_{\psi_i}p_{\psi_j} \,.
\eeqa
This is also an integrable Hamiltonian system. The following
functions are mutually commuting conserved quantities,
\beqa
F_1 &=& H,\nonumber\\
F_2^2&=&p_\chi^2+\frac{1}{ \sin^2\chi}
(p_{\eta}^2
-2(p_{\psi_2}+p_{\psi_1})p_\eta \cos\chi
+(p_{\psi_2}+p_{\psi_1})^2),\nonumber\\
F_3&=& p_{\eta},\qquad  F_4= p_{\psi_1}, \qquad F_5= p_{\psi_2}\,.
\eeqa
As in the homogeneous case, the new momentum coordinates are
introduced by
\beq
I_{a}(f)=\frac{1}{2\pi}\int_{C_a} \a \,.
\eeq
Four of the action variables are the same as in the homogeneous
case
\beq
|I_2|=f_2-\frac{1}{2}(|f_4+f_5+f_3|+|f_4+f_5-f_3|) \,,
\eeq
and
\beq
I_3=f_3\,,\quad I_4=\frac{2}{k_1}f_4\,, \quad
I_5=\frac{2}{k_2}f_5\,.
\eeq
The coordinate $I_1$ is harder to calculate. On the level set
\beq
M_f=\{(x,p) : F_a(x,p)=f_a \} \,,
\eeq
the z-component $p_z$ is written as
\beq\label{eq:pz}
p^2_z = \frac{h(z)^2}{4(1-z^2)} \left( f_1-
\frac{f_2^2-(f_4+f_5)^2}{b(z)^2}-a^{ij}(z) f_i f_j \right) \nonumber\\
\equiv \hat{Q}(z)\,,
\eeq
where
\beq
\hat{Q}(z)=\frac{Q_1}{(z-1)^2}+\frac{Q_2}{(z+1)^2}
+\frac{Q_3}{(z-z_0)^2} +\frac{R_1}{z-1}
+\frac{R_2}{z+1}+\frac{R_3}{z-z_0} \,,
\eeq
and the coefficients $Q_i, R_j$ are given by equations
(\ref{eq:Q}) and (\ref{eq:Q1}) and the equations in
appendix B, together with the replacements
\beq\label{eq:replace}
j(j+1) \rightarrow f_2^2, \quad  n_1, n_2 \rightarrow f_4, f_5,
\quad \lambda \rightarrow f_1\,.
\eeq

Given the expression (\ref{eq:pz}) for $p_z$, we would like to
calculate the action variable
\be
I_1 = \frac{1}{2\pi} \oint p_z dz \,.
\ee
Let us introduce
\beq
\varphi(z)=(z-1)^2(z+1)^2(z-z_0)^2 \hat{Q}(z) \,.
\eeq
The function $\varphi$ is a polynomial of degree 4 with
leading coefficient $-f_1/4$ and
\beq\label{eq:polyvals}
\varphi(1)=4 Q_1 (z_0-1)^2 \le 0, \quad \varphi(-1)=4 Q_2 (z_0+1)^2
\le 0, \quad \varphi(z_0)=Q_3 (z_0^2-1)^2 \le 0.
\eeq
The turning points of the geodesics are given by the roots of
$\varphi(z)$.
We will consider only the case where the polynomial
$\varphi$ has 4 distinct real roots, say
$\alpha_i$ $(i=1,2,3,4)$. Indeed, for large values
of $\nu_1$ and $\nu_2$, we can find real roots such that
either $(a) -1 < \alpha_1 < \alpha_2 < 1 < \alpha_3 < \alpha_4$, or
$(b) \, \alpha_1 < \alpha_2 < -1 <\alpha_3 < \alpha_4 < 1$,
depending on the values of $f_a$.
Thus the expression for the action variable becomes the following
elliptic integral
\beq\label{eq:action1}
I_1(f) = \frac{1}{\pi} \int_{\alpha_i}^{\alpha_j}
\frac{\sqrt{\varphi(z)}}{(z-1)(z+1)(z-z_0)} dz \,,
\eeq
where $\alpha_i$ and $\alpha_j$ are the roots between -1 and +1.

It seems that real roots $\alpha_i$ satisfying (a) or (b)
exist in the general case, although we have not proved this.
One can check that real roots $\alpha_i$ satisfying (a) or (b)
exist for small values of $f_4$, $f_5$ and any $\nu_i$.
Note that roots of the form
$-1<\alpha_1<\alpha_2<\alpha_3<1<\alpha_4$ or
$\alpha_1<-1<\alpha_2<\alpha_3<\alpha_4<1$ are forbidden by
(\ref{eq:polyvals}).

As a more explicit example, consider the rational limit $\nu_1,\nu_2
\to \infty$ with $\nu_2/\nu_1 = q > 1$ fixed. Keeping
$f_1,f_2,f_4,f_5$ fixed in the limit one finds $Q_i=0, (1=1,2,3)$ and
\beq
R_1=(1+q^2)\frac{f_2^2}{2}-\frac{f_1}{8},\quad
R_2=-(1+q^2)\frac{f_2^2}{2q^2}+\frac{f_1}{8},\quad
R_3=-R_1-R_2,
\eeq
This limit is given by (\ref{eq:rat1}) and (\ref{eq:rat2}) in appendix
B with $N_i=0$. Thus the polynomial becomes
\beqa
\varphi(z) &=& (z-1)^2(z+1)^2(z-z_0)^2 \hat{Q}(z) \nonumber\\
&\rightarrow& -\frac{f_1}{4}(z-1)(z+1)(z-z_0)(z-z_1),
\eeqa
where $z_0=(q^2+1)/(q^2-1)$ and $z_1=z_0(1-8 f_2^2/f_1)$.
Positivity of $\varphi$ and $|z|<1$ requires one of four cases
\beqa
(I) &-1<z<z_1<1<z_0,\quad (II) & -1<z<1<z_1<z_0,\nonumber\\
(III) &z_0<z_1<-1<z<1, \quad (IV) & z_0<-1<z_1<z<1.
\eeqa
Near the rational limit, the 4 real roots $ \{-1,1,z_0,z_1 \}$
change as
\beq
-1 \rightarrow -1+\delta_1, \quad 1 \rightarrow 1+\delta_2,
\quad z_0 \rightarrow z_0^{*}, \quad z_1 \rightarrow z_1^{*}\,,
\eeq
where $z_0^{*},z_1^{*}$ are the new values of $z_0,z_1$ and
$\delta_i$ are evaluated up to the order $\nu_i^{-2}$ as
\beq
\delta_1=-\frac{8Q_2(1+z_0)}{f_1(1+z_1)}, \quad
\delta_2= \frac{8Q_1(1-z_0)}{f_1(1-z_1)}\,.
\eeq
This implies the inequality
$(a) -1<\alpha_1<\alpha_2<1<\alpha_3<\alpha_4$ or
$(b) \, \alpha_1<\alpha_2<-1<\alpha_3<\alpha_4<1$
by $Q_i<0$ and $f_1>0$:
\beq
(I),(II) \rightarrow (a), \quad (III),(IV) \rightarrow (b).
\eeq
Therefore the action variable is of the form (\ref{eq:action1}). The
integral may be evaluated in the four cases

(I) $-1 < z < z_1 < 1 < z_0:$
\beq
|I_1|
= \frac{\sqrt{f_1}(1-z_1)}{ \pi \sqrt{2(z_0-z_1)}}
( \Pi(\pi/2, (z_1+1)/2, \kappa)-F(\pi/2, \kappa))\,,
\eeq
with
\beq
\kappa=\sqrt{\frac{(z_0-1)(z_1+1)}{2(z_0-z_1)}}.
\eeq

(II) $-1 < z < 1 < z_1 < z_0:$
\beq
|I_1|
= \frac{\sqrt{f_1}(z_1-1)}{ \pi \sqrt{(z_0-1)(z_1+1)}}
\Pi(\pi/2, 2/(z_1+1), \kappa)\,,
\eeq
with
\beq
\kappa=\sqrt{\frac{2(z_0-z_1)}{(z_0-1)(z_1+1)}}.
\eeq

(III) $ z_0  < z_1 < -1 < z < 1:$
\beq
|I_1|
= -\frac{\sqrt{f_1}(z_1+1)}{ \pi \sqrt{(z_0+1)(z_1-1)}}
\Pi(\pi/2, 2/(1-z_1), \kappa)\,,
\eeq
with
\beq
\kappa=\sqrt{\frac{2(z_1-z_0)}{(z_0+1)(z_1-1)}}.
\eeq

(IV) $z_0 < -1 < z_1 < z < 1:$
\beq
|I_1|
= \frac{\sqrt{f_1}(z_1+1)}{ \pi \sqrt{2(z_1-z_0)}}
( \Pi(\pi/2, (1-z_1)/2, \kappa)-F(\pi/2, \kappa))\,,
\eeq
with
\beq
\kappa=\sqrt{\frac{(z_0+1)(z_1-1)}{2(z_1-z_0)}}.
\eeq
In these expressions $F$ and $\Pi$ denote complete elliptic integrals
of the first and third kind respectively.

\subsection{Slightly inhomogeneous metrics: semiclassical spectrum}

In this subsection we calculate the action variable for a small
perturbation about the homogeneous metrics. This is the classical
computation corresponding to the spectral calculation of section 3.4
above.

The homogeneous metrics are given by $z_0 \rightarrow \infty$. In this
limit
\beq
\varphi(z) \rightarrow z_0^2 \,\hat{\varphi}(z)+o(z_0) \,,
\eeq
which represents the transition from the Riemann surface $y^2=\varphi(z)$ of genus 1 to
$y^2=\hat{\varphi}(z)$ of genus 0.
We therefore have
\beq
\hat{Q}(z)=\frac{\hat{\varphi}(z)}{(z-1)^2(z+1)^2},
\eeq
where $\hat{\varphi}$ is a polynomial of degree 2:
\beq
\hat{\varphi}(z)=Q_1(z+1)^2+Q_2(z-1)^2+(R_1-R_2)(z-1)(z+1).
\eeq
The coefficients $Q_i$ and $R_i$ are given by (\ref{eq:Q}),
(\ref{eq:Q1}) and (\ref{eq:homogcase}),
again with the replacements of (\ref{eq:replace})
\beq
j(j+1) \rightarrow f_2^2, \quad  n_1, n_2 \rightarrow f_4, f_5,
\quad \lambda \rightarrow f_1\,.
\eeq
The action variable is calculated using the integral (\ref{eq:area})
to give
\beq
|I_1(f)|=\sqrt{|Q_1|+|Q_2|-R_1+R_2}-\sqrt{|Q_1|}-\sqrt{|Q_2|}\,.
\eeq
Solving this expression for $f_1=H$ recovers the energy given in
(\ref{eq:actionhamil}).

Now consider a perturbation away from homogeneity as in section 3.4
\be
\frac{\nu_2}{\nu_1}= 1+\epsilon \,, \qquad\text{with}\qquad \epsilon
\ll 1\,.
\ee
In this case we have a polynomial of degree 3 by ignoring
the terms that are $o(\epsilon^2)$,
\beq\label{eq:polypert}
\hat{\varphi}(z)=Q_1(z+1)^2+Q_2(z-1)^2+(R_1-R_2)(z-1)(z+1)+
z(z-1)(z+1)(R_1+R_2).
\eeq
The coefficients $Q_i$ and $R_i$ are given by
(\ref{eq:Q}), (\ref{eq:Q1}) and (\ref{eq:coef1}), (\ref{eq:coef2})
with $\nu_2/\nu_1=1+\epsilon$. The last term in the polynomial
(\ref{eq:polypert}) has coefficient $R_1+R_2$ which is order $\epsilon$,
since $R_1+R_2=0$ for $\epsilon=0$. The integral for the action variable
can be evaluated as a perturbation of the homogeneous limit.
We find
\beqa
|I_1(f)|&=&\sqrt{|Q_1|+|Q_2|-R_1+R_2}-\sqrt{|Q_1|}-\sqrt{|Q_2|}
\nonumber\\
&-&\frac{(R_1+R_2)(Q_1-Q_2)}{2(|I_1|+\sqrt{|Q_1|}+\sqrt{|Q_2|})^3}
+o(\epsilon^2).
\eeqa

Expressing the coefficients in terms of the homogeneous quantities $I_4,I_5$, namely
$Q_1=-I_4^2/4+\epsilon Q_1^{(1)}$ and $Q_2=-I_5^2/4+\epsilon Q_2^{(1)}$,
together with (\ref{eq:Qpert}) and (\ref{eq:Rpert}) in appendix B,
we have
\beq\label{semiclass2}
f_1=f_1 \Bigl|_{\text{homogeneous}}+\epsilon
\frac{4(2+\nu^2)}{1+\nu^2}f_1^{(1)}+o(\epsilon^2)\,,
\eeq
where
\beqa\label{semiclass3}
f_1^{(1)}&=& Q_1^{(1)}+Q_2^{(1)}+R_1^{(1)}-R_2^{(1)}
-(2|I_1|+|I_4|+|I_5|)\left( \frac{Q_1^{(1)}}{|I_4|}+
\frac{Q_2^{(1)}}{|I_5|} \right) \nonumber\\
&-&\frac{(I_4^2-I_5^2)(R_1^{(1)}+R_2^{(1)})}{(2|I_1|+|I_4|+|I_5|)^2}.
\eeqa
The perturbed energy is just $H=f_1$. Therefore the semiclassical
spectrum is computed by setting $|I_1| \rightarrow L$,
$I_4 \rightarrow N_1$ and $I_5\rightarrow N_2$ in equations
(\ref{semiclass2}) and (\ref{semiclass3}). It is satisfying to see
that the resulting semiclassical spectrum agrees with the Laplacian
spectrum (\ref{eq:perturbed}) within a semi-classical
approximation. This provides a check on our calculations.
We can understand why the first term in (\ref{eq:perturbed}) is
lacking from the semiclassical result by again comparing a naive
quantisation of the classical Hamiltonian (\ref{eq:inhamil}), $p_a \to
i\pa/\pa x^a$, with the Laplacian (\ref{eq:laplacian}). There is an
ordering ambiguity, and we see that the
$P(z)d/dz$ term in the Fuchs equation does not exist in the
quantiation of the classical Hamiltonian.

\section{Conclusions and future directions}

We have studied the Laplacian spectrum, Lichnerowicz spectrum and
geodesics on an infinite family of five dimensional inhomogeneous
Einstein metrics. The moduli space of metrics included an infinite
sequence of homogeneous metrics, $T^{p,1}$. For the homogeneous
metrics, we were able to give very explicit results for the Laplacian
spectrum and for the frequences of closed geodesics. Further, we were
able to use the explicit spectrum on $T^{p,1}$ to calculate the zeta
function on these metrics at large $p$.

The inhomogeneous metrics are harder to study. We have shown that the
Laplace equation and Hamilton-Jacobi equation may be separated for
these cases. We found some perturbative results for the Laplacian
spectrum for slightly inhomogeneous spaces. However, it seems that the
full inhomogeneous spectrum will require numerical
calculations or more sophisticated methods than we have used.

For the geodesics in the inhomogeneous case we have identified four of
the five action variables. The calculation of the remaining action
variable reduces to an elliptic integral with an underlying elliptic
curve of genus one. If one could
calculate the remaining action variable, then it is possible that the
semiclassical quantisation of the system would give some insight into
the Laplacian spectrum for the inhomogeneous metrics.
Semiclassical quantisation of the homogeneous and slightly
inhomogeneous cases gives a good agreement with the Laplacian spectrum.

The Lichnerowicz spectrum contains information about the stability of
Freund-Rubin compactifications and generalised black holes constructed
from Einstein metrics. We saw that of the family of metrics we
considered, only $T^{1,1}$ can be shown to give a stable Freund-Rubin
compactification and generalised black holes with vanishing
cosmological constant. We suspect that the remaining spacetimes are
all unstable. Unstable generalised black holes are interesting as it
is unclear what the endpoint of the instability will be and the
instability may give rise to interesting dynamics, analogous to what has
been discovered recently for the black string instability
\cite{Choptuik:2003qd}.

We saw that the moduli space contains metrics with an
arbitrarily large maximum of Weyl eigenvalues.
This occurs in the rational limit at large inhomogeneity $q$, see
Table 1 above. This fact suggests that
the minimum Lichnerowicz eigenvalue may become arbitrarily negative
\cite{Gibbons:2002th}, giving rise to unstable generalised
Anti-de Sitter black holes. The instability of these black holes
predicts a thermal instability of a dual theory propagating on
the corresponding Einstein metric \cite{Hartnoll:2004kz,Hartnoll:2003as}.
If one could calculate the zeta function for these backgrounds, the
corresponding thermodynamics may help elucidate the
nature of the predicted field theory instability. Perhaps the
Laplacian spectrum in the large-$q$ rational limit is calculable?
One should note that it is not certain that the dual
instability will necessarily be present for a free scalar field, as
duality relates the $AdS_7$ black hole to the the strongly coupled
thermal theory living on $M5$ branes which is certainly much more
complicated. However, it seems possible that a thermodynamic
instability for field theories on a curved background
with regions of large curvature is a generic phenomenon.

Finally, it would be interesting to perform an analysis similar to ours
for other Einstein metrics. In particular, Einstein-Sasaki metrics
always give stable Freund-Rubin compactifications and stable
generalised black holes \cite{Gibbons:2002th}. Therefore a study of
the Laplacian spectrum of the five dimensional Einstein-Sasaki metrics
constructed in \cite{Gauntlett:2004yd} may have interesting
applications to the AdS/CFT correspondence. These metrics are also
cohomogeneity one and it seems clear that the equations can also be
separated in these cases.

\subsection*{Acknowledgements}

S.A.H. is supported by the Sims scholarship.
The research of Y.Y is supported by the 21 COE program
``Constitution of wide-angle mathematical basis focused on knots''
and the Grant-in Aid for scientific Research (No.14540073, 14540275)
from the Japanese Ministry of Education.

\appendix

\section{Metric terms}

We normalise the metrics so that the Ricci scalar is $R=20$.
\bea
h^2(\q) &=&
\frac{(1-\nu_1^2 \nu_2^2)}{(2-\nu_1^2 - \nu_2^2)}
\frac{1-\nu_1^2\cos^2\theta-\nu_2^2\sin^2\theta}
{1-\mu_1^2\cos^2\theta-\mu_2^2\sin^2\theta}\,,
\\
a_{11}(\q) &=&
\frac{\nu_1^2(1-\nu_2^2)^2(2-\nu_1^2-\nu_2^2)(1-\nu_1^2 \nu_2^2)}
{4(1+\nu_1^4\nu_2^2+\nu_1^2\nu_2^4-3\nu_1^2\nu_2^2)^2}
\frac{(1-\mu_1^2\cos^2\theta-\nu_2^2\sin^2\theta)\sin^2\theta}
{1-\nu_1^2\cos^2\theta-\nu_2^2\sin^2\theta}\,,
\\
a_{22}(\q) &=&
\frac{\nu_2^2(1-\nu_1^2)^2(2-\nu_1^2-\nu_2^2)(1-\nu_1^2 \nu_2^2)}
{4(1+\nu_1^4\nu_2^2+\nu_1^2\nu_2^4-3\nu_1^2\nu_2^2)^2}
\frac{(1-\nu_1^2\cos^2\theta-\mu_2^2\sin^2\theta)\cos^2\theta}
{1-\nu_1^2\cos^2\theta-\nu_2^2\sin^2\theta}\,,
\\
a_{12}(\q) &=&
-\frac{\nu_1^2\nu_2^2(1-\nu_1^2)^2(1-\nu_2^2)^2(1-\nu_1^2 \nu_2^2)}
{4(1+\nu_1^4\nu_2^2+\nu_1^2\nu_2^4-3\nu_1^2\nu_2^2)^2}
\frac{\sin^2\theta\cos^2\theta}{1-\nu_1^2\cos^2\theta-\nu_2^2\sin^2\theta}\,,
\\
b^2(\q) &=&
\frac{(1-\nu_1^2 \nu_2^2)(1-\nu_1^2\cos^2\theta-\nu_2^2\sin^2\theta)}
{4(1+\nu_1^4\nu_2^2+\nu_1^2\nu_2^4-3\nu_1^2\nu_2^2)} \,,
\end{eqnarray}
with
\begin{eqnarray}
\mu_1^2=\frac{\nu_1^2(1-\nu_1^2\nu_2^2)}{2-\nu_1^2-\nu_2^2}\,,~~~~
\mu_2^2=\frac{\nu_2^2(1-\nu_1^2\nu_2^2)}{2-\nu_1^2-\nu_2^2}\,.
\eea
The metric has the following integrality conditions
\bea
k_1&=&
\frac{\nu_1(1-\nu_2^2)(2-\nu_2^2-\nu_1^2\nu_2^2)}
{1+\nu_1^4\nu_2^2+\nu_1^2\nu_2^4-3\nu_1^2\nu_2^2} \in \Z \,,
\\
k_2&=&
\frac{\nu_2(1-\nu_1^2)(2-\nu_1^2-\nu_1^2\nu_2^2)}
{1+\nu_1^4\nu_2^2+\nu_1^2\nu_2^4-3\nu_1^2\nu_2^2} \in \Z \,.
\eea
In the homogeneous case, $\nu_1 = \nu_2 = \nu$,
\bea
a^{11} & = & \frac{2 (2\nu^2+1)^2}{(\nu^2+2)(1+\nu^2)} \left(1 +
\frac{2}{\nu^2 \sin^2\q} \right) \,, \nonumber \\
a^{22} & = & \frac{2 (2\nu^2+1)^2}{(\nu^2+2)(1+\nu^2)} \left(1 +
\frac{2}{\nu^2 \cos^2\q} \right) \,, \nonumber \\
a^{12} & = & \frac{2 (2\nu^2+1)^2}{(\nu^2+2)(1+\nu^2)} \,, \nonumber \\
b^{-2} & = & \frac{4 (2\nu^2+1)}{1+\nu^2} \,,\quad h^{-2} =
\frac{\nu^2+2}{1+\nu^2} \,, \quad \sqrt{\det a}  \propto  \sin\q \cos\q \,.
\eea

\section{Coefficients in the inhomogeneous equation}

The coefficients $R_i$ are given by
\beqa\label{eq:coef1}
R_1&=&\frac{(j(j+1)-(n_1+n_2)^2)B}{8(1-\mu_1^2)}
-\frac{A_1 r_{+}n_1^2}{2}
+\frac{A_2 (1-\nu_1^2)n_2^2}{8(1-\mu_1^2)}
+\frac{A_{12}(1-\nu_1^2)n_1 n_2}{4(1-\mu_1^2)^2}\nonumber\\
&-&\frac{(1-\nu_1^2)(1-\nu_1^2 \nu_2^2)\lambda}
{8(2-\nu_1^2-\nu_2^2)(1-\mu_1^2)} \,,
\eeqa
\beqa\label{eq:coef2}
R_2&=&-\frac{(j(j+1)-(n_1+n_2)^2)B}{8(1-\mu_2^2)}
+\frac{A_2 r_{-}n_2^2}{2}
-\frac{A_1 (1-\nu_2^2)n_1^2}{8(1-\mu_2^2)}
-\frac{A_{12}(1-\nu_2^2)n_1 n_2}{4(1-\mu_2^2)^2}\nonumber\\
&+&\frac{(1-\nu_2^2)(1-\nu_1^2 \nu_2^2)\lambda}
{8(2-\nu_1^2-\nu_2^2)(1-\mu_2^2)} \,, \nonumber \\
R_3 & = & -R_1 - R_2 \,,
\eeqa
where
\beqa
A_1 &=& \frac{4(1+\nu_1^4 \nu_2^2+\nu_1^2 \nu_2^4-3 \nu_1^2 \nu_2^2)^2}
{\nu_1^2(1-\nu_2^2)^2(2-\nu_1^2-\nu_2^2)^2} \,,\nonumber\\
A_2 &=&  \frac{4(1+\nu_1^4 \nu_2^2+\nu_1^2 \nu_2^4-3 \nu_1^2 \nu_2^2)^2}
{\nu_2^2(1-\nu_1^2)^2(2-\nu_1^2-\nu_2^2)^2} \,,\nonumber\\
A_{12} &=& \frac{4(1+\nu_1^4 \nu_2^2+\nu_1^2 \nu_2^4-3 \nu_1^2 \nu_2^2)^2}
{(2-\nu_1^2-\nu_2^2)^3} \,, \nonumber\\
B &=& \frac{4(1+\nu_1^4 \nu_2^2+\nu_1^2 \nu_2^4-3 \nu_1^2 \nu_2^2)}
{2-\nu_1^2-\nu_2^2} \,,
\eeqa
and
\beqa
r_{+} &=& -\frac{c_1(1-\nu_1^2)^2}{(1-\mu_1^2)^3}
+\frac{c_2(1-\nu_1^2)}{2(1-\mu_1^2)^2}
+\frac{c_3(1-\nu_1^2)}{2(1-\mu_1^2)^2}
-\frac{(1-\nu_1^2)^2}{4(1-\mu_1^2)^2} \,,\nonumber\\
r_{-} &=& \frac{c_1(1-\nu_2^2)^2}{(1-\mu_2^2)^3}
-\frac{c_2(1-\nu_2^2)}{2(1-\mu_2^2)^2}
-\frac{c_4(1-\nu_2^2)}{2(1-\mu_2^2)^2}
-\frac{(1-\nu_2^2)^2}{4(1-\mu_2^2)^2} \,,
\eeqa
with
\beqa
c_1 &=& -\frac{(\nu_1^2-\nu_2^2)(1-\nu_1^2 \nu_2^2)}
{2(1-\nu_1^2-\nu_2^2)} \,,\nonumber\\
c_2 &=& \frac{-\nu_1^2+\nu_2^2}{2} \,,\nonumber\\
c_3 &=&
\frac{-2 \nu_1^2+\nu_2^2+\nu_1^4+\nu_1^2 \nu_2^2-\nu_1^2 \nu_2^4}
{2(2-\nu_1^2-\nu_2^2)} \,,\nonumber\\
c_4 &=&
\frac{2 \nu_2^2-\nu_1^2-\nu_2^4-\nu_1^2 \nu_2^2+\nu_1^4 \nu_2^2}
{2(2-\nu_1^2-\nu_2^2)}\,.
\eeqa
In the homogeneous case, $\nu_1=\nu_2=\nu$, we have
\beqa\label{eq:homogcase}
R_1 &=& -R_2 \nonumber\\
&=& \frac{(1+2 \nu^2)(j(j+1)-(n_1+n_2)^2)}{2(2+\nu^2)}
+\frac{(1+2 \nu^2)^2(n_1^2+n_2^2)}{4 \nu^2(2+\nu^2)}
+\frac{(1+2 \nu^2)^2 n_1 n_2}{2(2+\nu^2)^2}
-\frac{(1+\nu^2)\lambda}{8(2+\nu^2)} \,,\nonumber\\
R_3 &=& 0 \,.
\eeqa

We also collect here the terms for the perturbation about the
homogeneous metrics. The required terms for $Q^{(1)}$ are
\beqa\label{eq:Qpert}
Q_1^{(1)} & = & \frac{4 (2\nu^2+1)(\nu^4+2\nu^2+3)}{(\nu^2+2)^3(\nu^2-1)} n_1^2 \,, \nonumber \\
Q_2^{(1)} & = & - \frac{2
  (2\nu^2+1)(7\nu^4+3\nu^2+2)}{(\nu^2+2)^3 (\nu^2-1)\nu^2} n_2^2 \,,
\eeqa
whilst the terms for $R^{(1)}$ are
\beqa\label{eq:Rpert}
R_1^{(1)} &=& \frac{\nu^2( -1+3 \nu^2+\nu^4)}{(\nu^2-1)(\nu^2+2)^2}
(j(j+1)-(n_1+n_2)^2)
-\frac{20+56 \nu^2+37 \nu^4+8 \nu^6-4 \nu^8}{4(\nu^2-1)(\nu^2+2)^3}
n_1^2 \nonumber\\
&+& \frac{4+10 \nu^2+11 \nu^4+16 \nu^6+4 \nu^8}
{4\nu^2(\nu^2-1)(\nu^2+2)^2}n_2^2+
\frac{\nu^2(-6-5 \nu^2+16 \nu^4+4 \nu^6)}{2(\nu^2-1)(\nu^2+2)^3}
n_1 n_2 \nonumber\\
&+&\frac{\nu^2}{4(\nu^2-1)(\nu^2+2)^2}\lambda^{(0)} \,, \nonumber\\
R_2^{(1)} &=& \frac{\nu^2( 2+ \nu^4)}{(\nu^2-1)(\nu^2+2)^2}
(j(j+1)-(n_1+n_2)^2)
+\frac{10+23 \nu^2+8 \nu^4+4 \nu^6}{4(\nu^2-1)(\nu^2+2)^2}
n_1^2 \nonumber\\
&-& \frac{8+24 \nu^2+32 \nu^4+41 \nu^6+16 \nu^8-4 \nu^{10}}
{4\nu^2(\nu^2-1)(\nu^2+2)^3}n_2^2+
\frac{\nu^2(6+7 \nu^2-8 \nu^4+4 \nu^6)}{2(\nu^2-1)(\nu^2+2)^3}
n_1 n_2 \nonumber\\
&+&\frac{\nu^4}{4(\nu^2-1)(\nu^2+2)^2}\lambda^{(0)} \,.
\eeqa

Finally, the coefficients simplify in the
rational limit, $\nu_1,\nu_2 \to \infty$ with $\nu_2/\nu_1
= q > 1$ fixed. One obtains
\be
z_0= \frac{q^2+1}{q^2-1} \,.
\ee
The coefficients (\ref{eq:coeffs}) become, introducing the integers
$N_i$ by $n_i = k_i N_i/2$,
\beq\label{eq:rat1}
Q_1 = -\frac{(1+q^2)^2}{16 q^4} N_1^2\,,\quad Q_2 =
-\frac{(1+q^2)^2}{16 q^2} N_2^2\,, \quad Q_3 = -\frac{1+q^2}{16 q^2} (N_1+N_2)^2.
\eeq
Furthermore,
\beqa\label{eq:rat2}
R_1 &=& (1+q^2)\frac{j(j+1)}{2}-\frac{\lambda}{8} \nonumber\\
&+&\frac{1}{32 q^4}
(1+3 q^2-q^4-2 q^6)N_1^2
+ \frac{q^2}{32}N_2^2
-\frac{2+q^2}{16}N_1 N_2, \nonumber\\
R_2 &=& -(1+q^2)\frac{j(j+1)}{2 q^2}+\frac{\lambda}{8}
 \nonumber\\
&-& \frac{1}{32 q^4}N_1^2
+\frac{1}{32 q^4}(2+q^2-3 q^4-q^6)N_2^2
+\frac{1}{16 q^4}(1+2 q^2)N_1 N_2, \nonumber\\
R_3 &=& -R_1-R_2 .
\eeqa

\section{Properties of Jacobi polynomials}

All the following identities involving Jacobi polynomials may be
found, for example, in \cite{Jacob} section 8.96.

The first equation we use allows us to express $z P^{(\a,\b)}_n(z)$ in
terms of $P^{(\a,\b)}_n(z)$, $P^{(\a,\b)}_{n-1}(z)$ and
$P^{(\a,\b)}_{n+1}(z)$
\bea
\lefteqn{ 2(n+1)(n+\a+\b+1)(2n+\a+\b)P^{(\a,\b)}_{n+1}(z) \nonumber} \\
 & & = (2n+\a+\b+1)[(2n+\a+\b)(2n+\a+\b+2)z+\a^2-\b^2] P^{(\a,\b)}_n(z)
\nonumber \\
 & & - 2(n+\a)(n+\b)(2n+\a+\b+2)P^{(\a,\b)}_{n-1}(z) \,.
\eea

The following equation allows us to express $(1-z^2)
dP^{(\a,\b)}_n(z)/dz$ in terms of $P^{(\a,\b)}_n(z)$, $z
P^{(\a,\b)}_n(z)$ and $P^{(\a,\b)}_{n-1}(z)$
\bea
\lefteqn{(2n+\a+\b)(1-z^2) \frac{d}{dz} P^{(\a,\b)}_n(z) \nonumber} \\
 & & = n[(\a-\b)-(2n+\a+\b)z] P^{(\a,\b)}_n(z) +
2(n+\a)(n+\b)P^{(\a,\b)}_{n-1}(z) \,.
\eea

Once we have used the above relations, we use two integration
results. The first is
\bea
\lefteqn{\int_{-1}^{1} (1-z)^\a (1+z)^\b P^{(\a,\b)}_m(z)
  P^{(\a,\b)}_n(z) dz \nonumber} \\
 & & = 0 \,, \qquad \qquad \qquad \qquad \qquad \qquad \qquad \qquad\quad\,\, \text{if} \qquad m \neq n \nonumber \\
 & & = \frac{2^{\a+\b+1}\G(\a+n+1)\G(\b+n+1)}{n!(\a+\b+1+2n)\G(\a+\b+n+1)}
\,, \qquad \text{if} \qquad m=n \,.
\eea
The second relation we will use is
\be
\int_{-1}^{1} (1-z)^{\a-1} (1+z)^\b P^{(\a,\b)}_n(z)
  P^{(\a,\b)}_n(z) dz = \frac{2^{\a+\b}\G(\a+n+1)\G(\b+n+1)}{n!\,\a\,\G(\a+\b+n+1)} \,.
\ee

\section{Check of the zeta function}

The zeta function on an $n$ dimensional compact manifold (in our
case $n=5$),
\be
\zeta (s)=\sum_{i;\lambda_{i}>0} \lambda_{i}^{-s} \,,
\ee
can be shown to converge absolutely in the region $\Re s > n/2$,
and can be analytically extended to a meromorphic function of $s$
in the whole complex plane. The poles are located at
$s=n/2-k,\quad (k=0,1,2,\cdots$) and their residues are given by
\beq
\text{Res}_{s=n/2-k} = \frac{a_{2k}}{(4 \pi)^{n/2} \Gamma(n/2-k)} \,.
\eeq
If the operator is $A = -\square+\kappa$ then the first few $a_k$ are given by
\beq
a_0=\text{vol}(M),\quad a_2=- a_0 \,\kappa+\frac{1}{6}\int_M dx R
\,,
\eeq
and
\beq
a_4=\int_M dx
(\frac{1}{180}R^{abcd}R_{abcd}-\frac{1}{180}R^{ab}R_{ab}+
\frac{1}{72} R^2-\frac{1}{30}\nabla_a \nabla^a R) \,.
\eeq

Is our computation of the homogeneous zeta function able to
reproduce any of these results? We have just recalled that there
should be a pole at $s=5/2$ with residue
\be
\text{Res}_{s=5/2} = \frac{\text{vol}(M)}{(4\pi)^{5/2}\G(5/2)} \,.
\ee
In the large $\nu$ limit, this gives
\be
\text{Res}_{s=5/2} = \frac{\sqrt{2}}{24 \nu} \,.
\ee
We can compare this with our results. We have already noted that
case I is always subleading compared to cases III and IV. We see
that cases III and IV do have a pole at $s=5/2$. The total residue
of $\zeta(s)$ in equation (\ref{eq:thezeta}) turns out to be
\be
\text{Res}_{s=5/2} = \frac{\sqrt{2}}{24 \nu} \,.
\ee
Which is precisely as required! We see that case II cannot
contribute to this expression because it has no terms that are
${\mathcal{O}}(1/\nu)$. Thus we have a rather nontrivial check on
our calculation.

It is not possible to check the other residues because we have
only worked to leading order in $\nu$. The other poles do not
appear to this order.

\section{Heun's equation}

In order to  transform the Fuchsian equation of section 3.3, equations
(\ref{eq:form}) to (\ref{eq:coeffs}), to a canonical form write
\beq
\phi(z)=(1-z)^{\sqrt{|Q_1|}}(1+z)^{\sqrt{|Q_2|}}(z-z_0)^{\sqrt{|Q_3|}}
f(z)\,.
\eeq
The new variable $f(x)$ with  $x=(1-z)/2$
satisfies Heun's equation
\beq
\frac{d^2}{dx^2}f+\left(
\frac{\gamma}{x}+\frac{\delta}{x-1}+
\frac{\epsilon}{x-x_0} \right)\frac{d}{dx}f
+\frac{\alpha \beta x-q}{x(x-1)(x-x_0)}f=0,
\eeq
where
\beqa
\alpha&+&\beta=2\left(1+\sum_{i=1}^{3}\sqrt{|Q_i|} \right),\nonumber\\
\alpha \beta &=& 2\left(\sum_{i=1}^{3}\sqrt{|Q_i|}
+\sum_{i<j}\sqrt{|Q_iQ_j|}+(R_1+R_2)x_0-R_2  \right), \nonumber \\
\gamma &=& 1+2 \sqrt{|Q_1|}, \nonumber \\
\delta &=& 1+2 \sqrt{|Q_2|}, \nonumber \\
\epsilon &=& 1+2 \sqrt{|Q_3|}, \nonumber \\
q &=& (\sqrt{|Q_1|}+\sqrt{|Q_2|}+2\sqrt{|Q_1 Q_2|})x_0+
\sqrt{|Q_1|}+\sqrt{|Q_3|}+2\sqrt{|Q_1 Q_3|}+2 R_1 x_0.
\eeqa
One has
\beq
1+\alpha+\beta-\gamma-\delta-\epsilon = 0 \,.
\eeq

It is known that Huen's equation admits an expression in terms of
elliptic functions and this expression is closely related to
the Inozemtsev system. For example see \cite{SS,TT}. From these
references, if all of $\gamma, \delta, \epsilon$ are half-odd-integer
one can obtain exact solutions of Huen's equation.
Unfortunately in our case this condition is not satisfied:
$\gamma, \delta$ are integers and $\epsilon \in \R$.

\end{document}